\documentstyle[12pt]{article}
\begin{document}

\centerline{\Large\bf Cascades in the Earth's magnetosphere}
\vskip 0.6truecm
\centerline{\Large\bf initiated by photons with the parameters}
\vskip 0.6truecm
\centerline{\Large\bf of the highest energy AGASA events}

\vskip 1.5truecm
\centerline{\large W\l odzimierz Bednarek}

\vskip 0.7truecm
\centerline{Department of Experimental Physics, University of \L \'od\'z,}
\centerline{90-236 \L \'od\'z, ul. Pomorska 149/153, Poland}

\vskip 1.truecm

\centerline{Abstract}

\vskip 0.7truecm
We investigate the cascading effects of extremely high energy (EHE)
photons in the Earth's magnetosphere 
assuming that these photons arrive with the parameters of the highest energy AGASA
events (energies, arrival directions). For the location of the AGASA 
Observatory, we determine the directions in the
sky from which photons can cascade with a high (low) probability. In the case
of the primary photons with the parameters of the events with the energies $>
10^{20}$ eV,  we compute the average cascade  spectra
of secondary photons entering the Earth's atmosphere, and estimate their
fluctuations around these average values by selecting the events with the largest 
and smallest number of secondary cascade photons.
It is shown that most photons with the parameters of the highest energy AGASA 
events should initiate cascades in the Earth's 
magnetosphere with a high probability even though they tend to arrive from 
directions in the sky for which the perpendicular component of the magnetic 
field is weaker. On the other hand, if these events are caused by the photons with 
lower energies, then the fluctuations in their shower development in the
magnetosphere and the atmosphere should be higher than in the case of photons
with the energies estimated by the AGASA experiment.

\section{Introduction}

It is believed that the number of photons in the cosmic rays with the energies 
above $10^{19}$ eV may be significant. Extremely high
energy cosmic ray hadrons should produce many photons in collisions
with the Microwave Background Radiation~ when propagating through the 
universe (Wdowczyk \& Wolfendale~1990, Halzen
et al.~1990). The EHE photons may also be efficiently produced from the decay of
massive  particles (e.g. Higgs and gauge bosons), as predicted by some more 
exotic theories (see for review Bhattacharjee \& Sigl~2000). Therefore,
the investigation of the photon content in the EHE cosmic rays (CR) can give us
important information about its origin. It is very difficult to obtain experimental 
constraints on the content of EHE CR due to the low statistics of detected particles 
at these energies and problems with the 
particle interaction models. In spite of these problems first efforts have been 
undertaken. For example,  
the analysis of the Haverah Park data allows the authors to put the upper 
limit on the photonic content in the cosmic rays above $4\times 10^{19}$ eV equal to 
55\% (Ave et al.~1999, 2002). Similar limits have recently been obtained from the 
analysis of the AGASA data (Shinozaki et al.~2002). On the other hand, 
there is also
experimental evidence, e.g. small scale clustering of EHE CR with 
the energies above $4\times 10^{19}$ eV (Takeda et al.~1999, 2001, 
Tinyakov \& Tkachev~2001), which suggest that the EHE cosmic rays contain a 
significant part of neutral particles ($\gamma$-rays, neutrons ?).

Even if photons may not completely dominate the highest energy cosmic rays, it 
is useful to investigate the hypothesis of their photonic origin since
photons of sufficiently high 
energies may convert into $e^\pm$ pairs when propagating in a 
perpendicular magnetic field (Erber~1966). It has been shown by McBreen \& 
Lambert~(1981) that such a conversion of photons 
should also occur in the Earth's magnetic field.
The secondary $e^\pm$ pairs created by photons should
produce many secondary synchrotron photons, which in turn can again
convert into $e^\pm$ pairs. Cascades initiated by the primary photons
in the Earth's magnetosphere have already been discussed in the general
context of the photon initiated cascade theory in the Earth's
atmosphere by e.g. Aharonian, Kanevsky \& Sahakian (1991), Karaku\l a (1997),
Kasahara (1997), Stanev \& Vankov (1997), Anguelov \& Vankov (2000) and also in the 
context of detecting high energy photons by planned Auger
Observatories by Bednarek (1999) and Bertou, Billoir \& Dagoret-Campagne
(2000). 

Different experiments (Volcano Ranch, Haverah Park, Yakutsk, Fly's Eye,
AGASA) have reported several showers with the energies $> 10^{20}$eV 
(Nagano \& Watson~2000, Takeda et al.~1999, Hayashida et al.~2000). 
Best documented are the events observed by the AGASA 
array\footnote{available through AGASA web page: 
www-akeno.icrr.u-tokyo.ac.jp/AGASA/pub/}. This experiment registered
64 events with the energies $> 4\times 10^{19}$ eV (between them 
10 have energies above $10^{20}$ eV) with zenith angles smaller than 45$^{\rm o}$.
Extending the analysis of the AGASA data to the events coming for the zenith 
angles up to 60$^{\rm o}$ 
increased the number of candidate events above $10^{20}$ eV 
to 17 (Sakaki et al.~2001). 
Unfortunately, the information which allow us to determine of the azimuth 
and zenith angles of these events, has only been published for 47 events with 
the energies $>4\times 10^{19}$ eV, 8 of them have the energies $> 10^{20}$ eV 
(Hayashida~2000). 
The parameters of two additional recent events with the energies above $10^{20}$ eV
are reported on the AGASA web page.

In this paper we analyse the features of cascades initiated 
by photons assuming that they arrive in the Earth's magnetosphere with
the arrival directions of the AGASA events and energies 
$> 5\times 10^{19}$ eV. Our purpose is to find out if specific photons with
the parameters of the AGASA events may interact with the Earth's magnetic field.

\section{Cascades initiated by photons in the Earth's magnetic field}

The EHE photon with the energy $E_\gamma$ can convert into an $e^\pm$ pair 
in the magnetic field $B$ if the value of the parameter  
$\chi_\gamma = (E_\gamma/2mc^2)(B/B_{cr})$ 
(where $B_{\rm cr}\approx 4.414\times 10^{13}$G and $mc^2$ is the electron
rest mass) is high enough (Erber 1966). The secondary $e^\pm$ 
pairs can in turn produce synchrotron photons in the magnetic field.
The efficiency of this process is determined by the parameter 
$\chi_{e^\pm} = (E_{\rm e}/mc^2)(B/B_{cr})$. 
The energies of synchrotron photons can be high enough to produce
the next generation of $e^\pm$ pairs, sustaining, in this way, the development of 
a cascade in the magnetic field. The simulations of the primary photons with 
energies of the highest energy cosmic ray events entering the Earth's magnetosphere 
show that up to a few generations of secondary $e^\pm$ pairs can be produced. 

\begin{figure} 
\vspace{19.truecm} 
\includegraphics{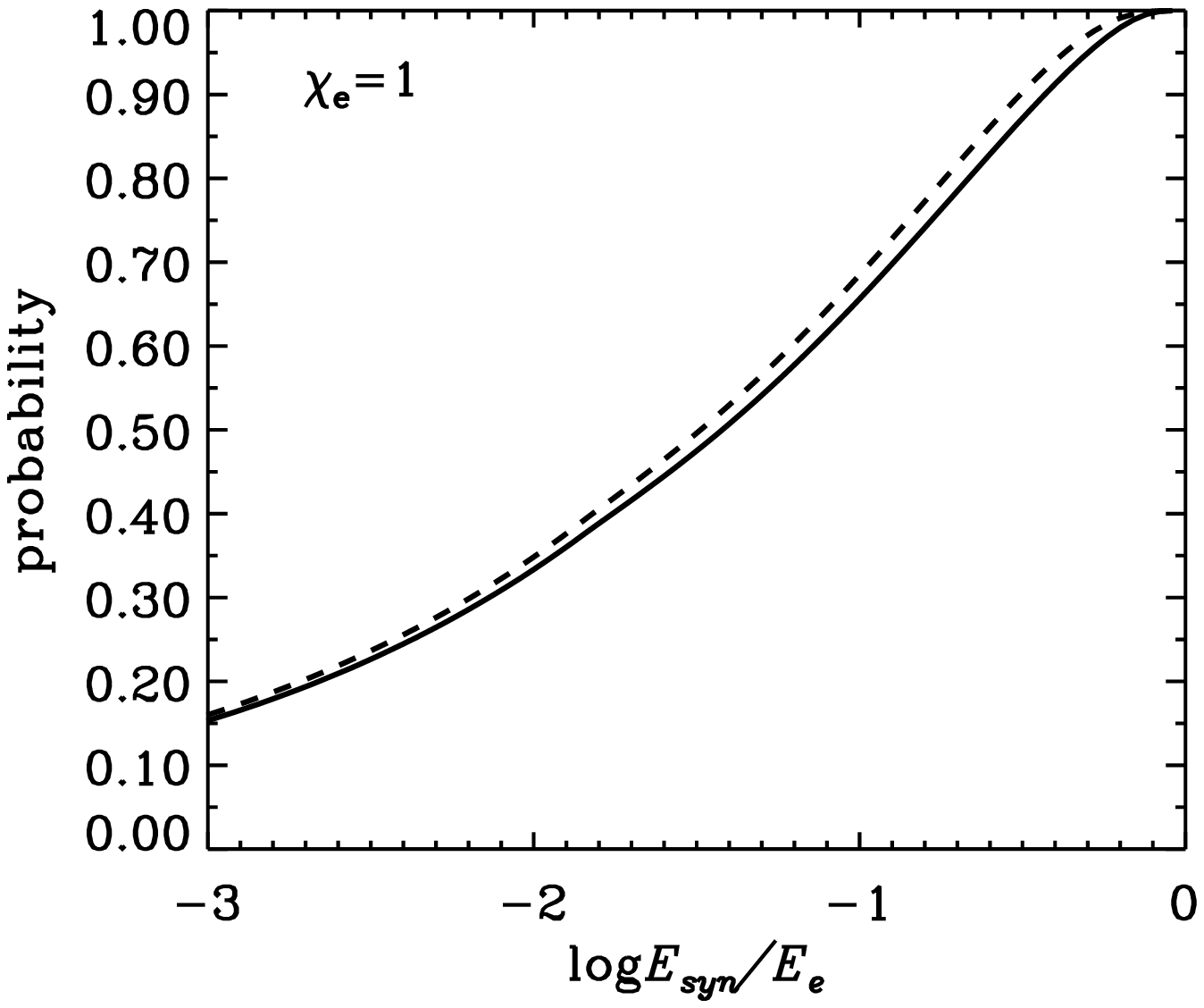}
\includegraphics{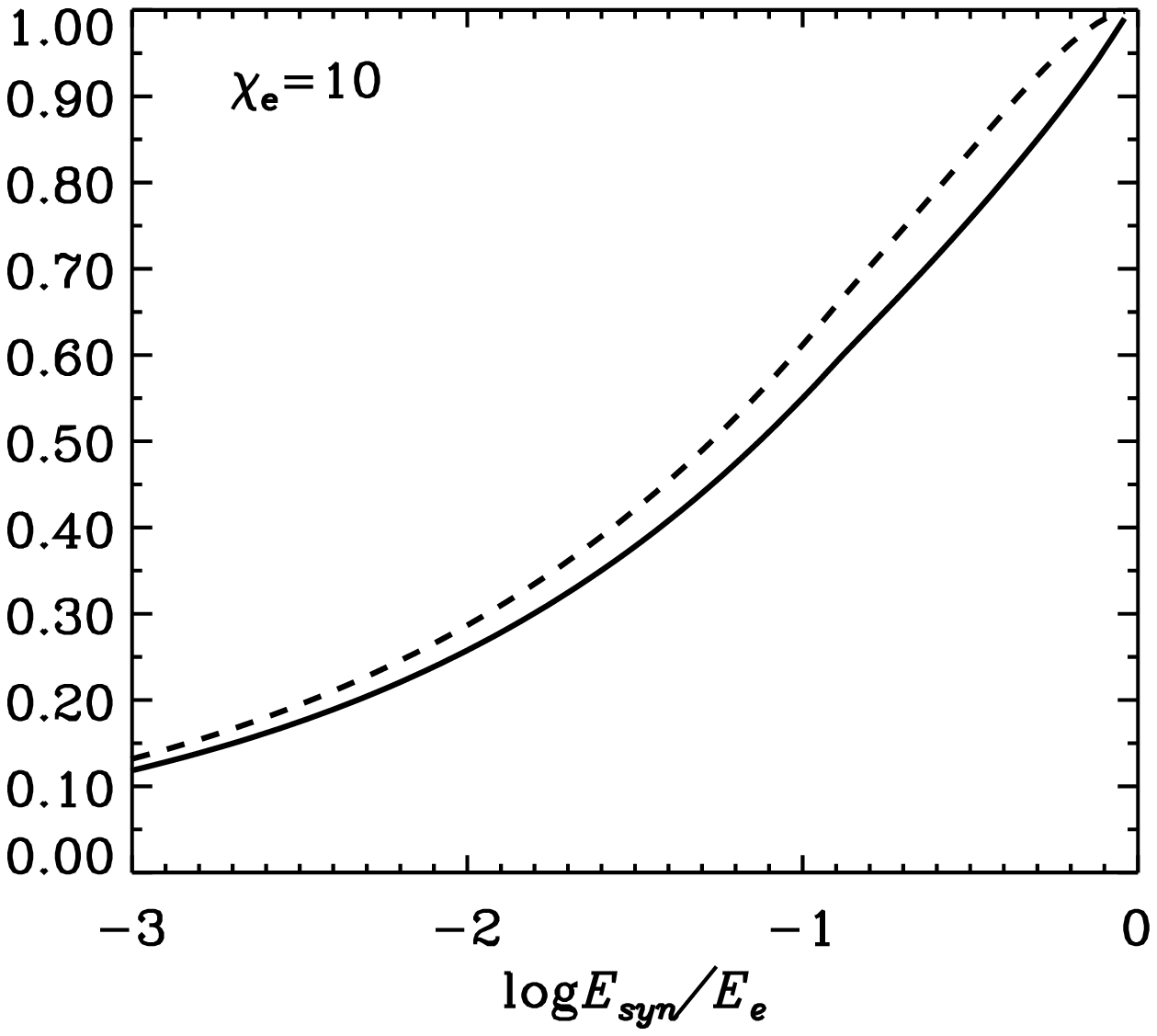}
\includegraphics{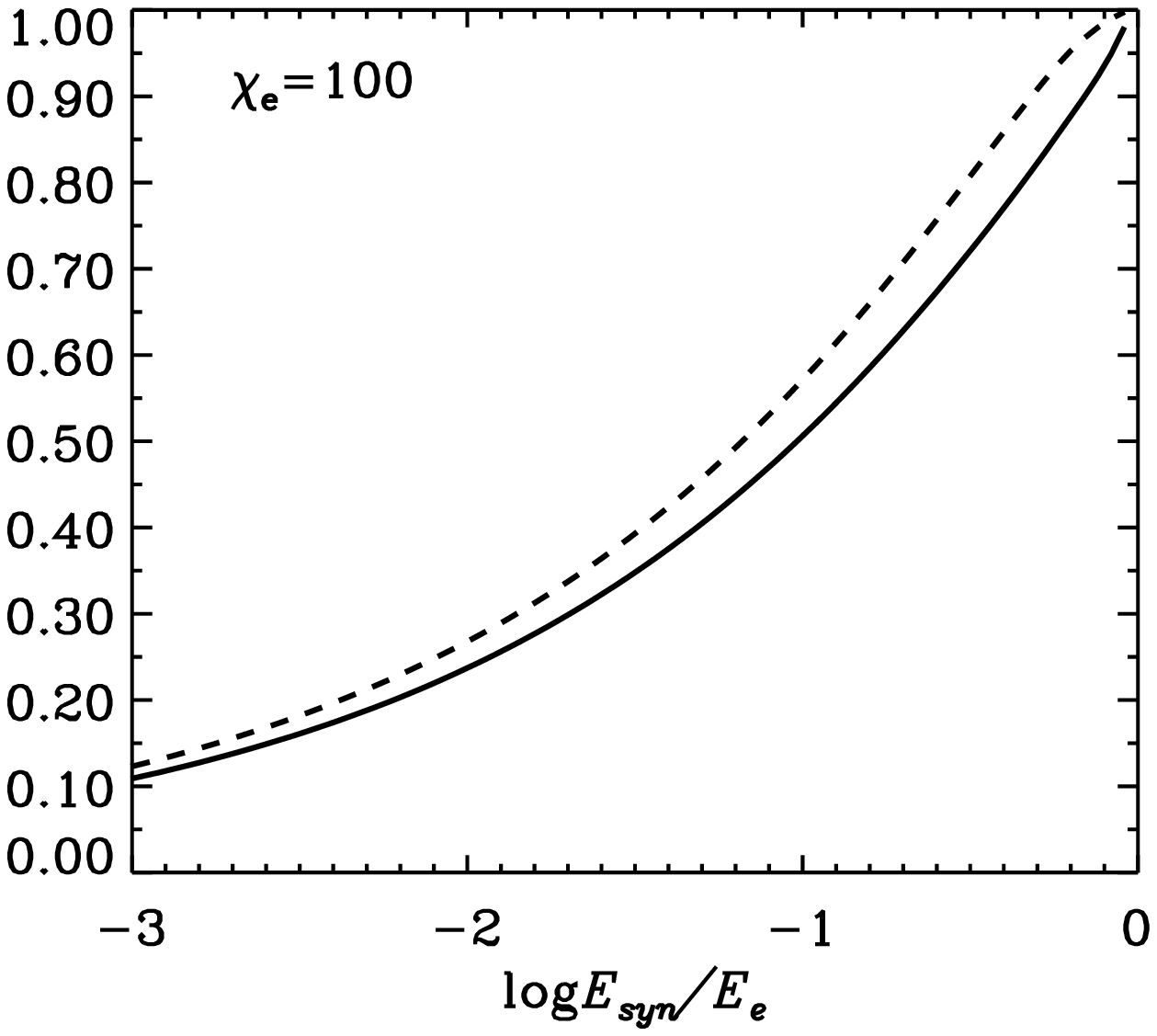}
\includegraphics{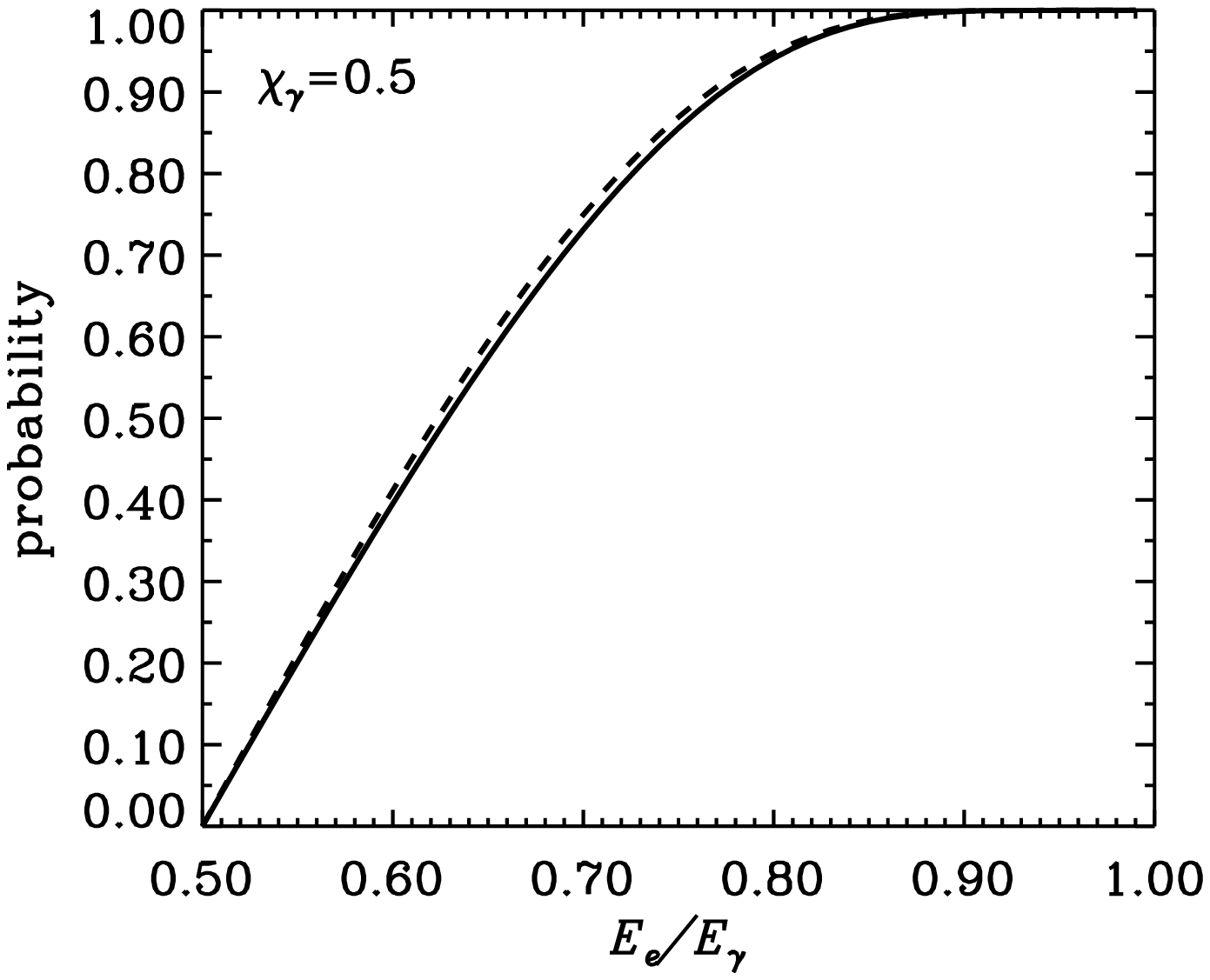}
\includegraphics{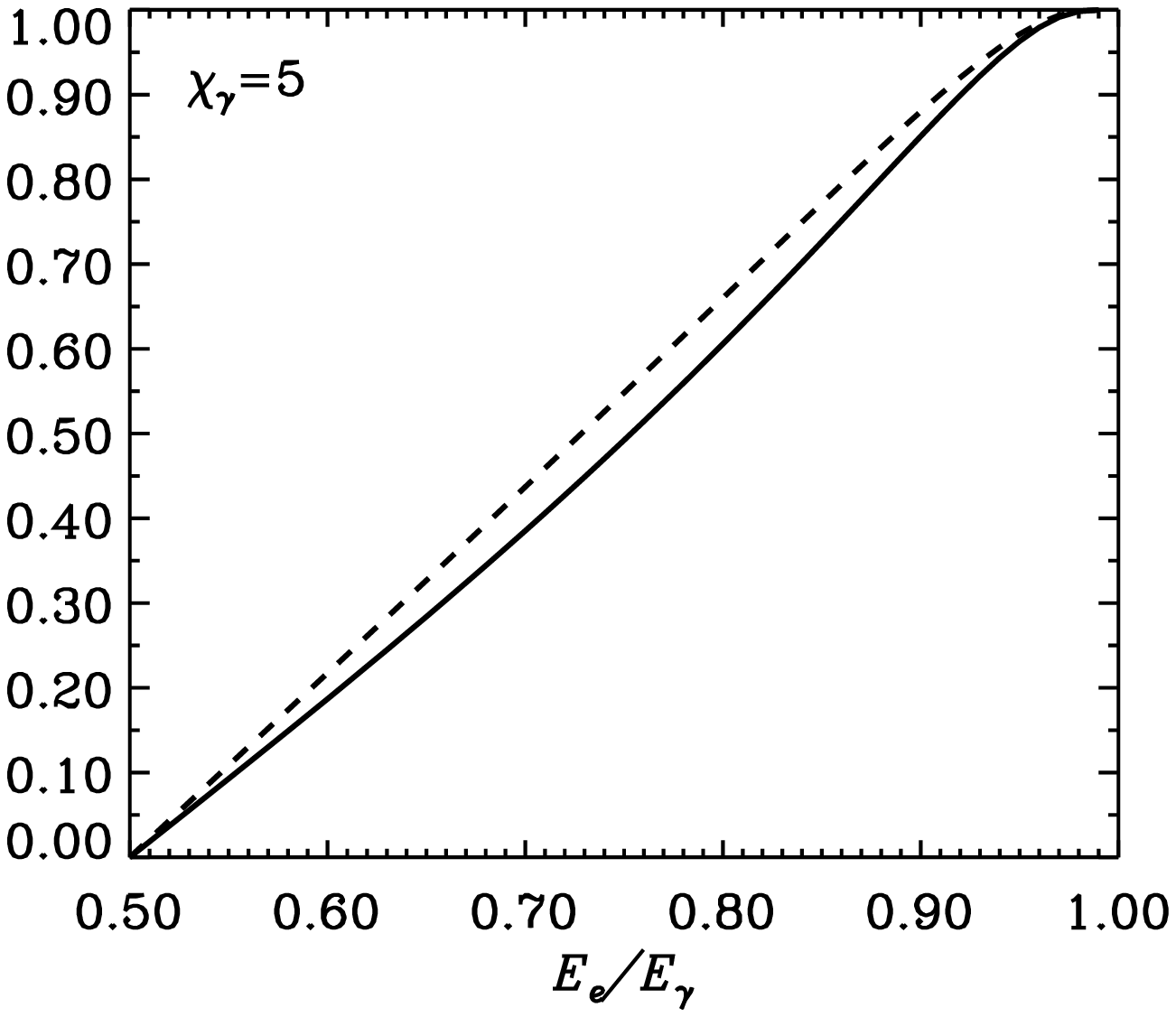}
\includegraphics{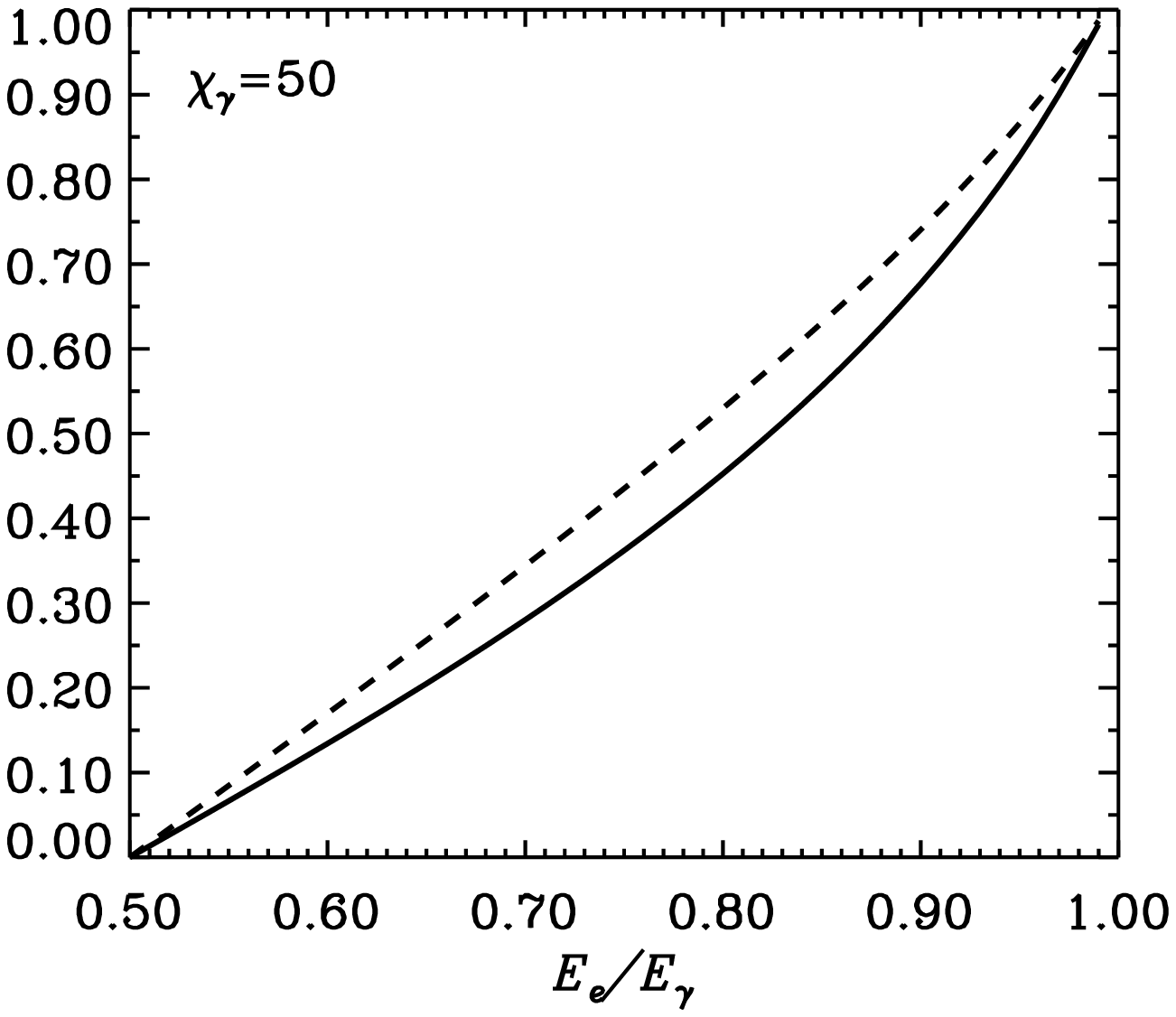}         
\caption[]{The probabilities of production of an $e^\pm$ pair
with energy $E_{\rm e}$ by the photon with energy $E_\gamma$
(top figures) and for emission of synchrotron photons with energy
$E_{\rm syn}$ by the electron with energy $E_{\rm e}$ (bottom 
figures) for the selected values of the 
parameters $\chi_\gamma$ and  $\chi_{\rm e}$. The computations with the
Baring's rates are shown by the full curves and with the Erber's 
rates by the dashed curves.}   
\label{fig0} 
\end{figure}

We simulate the development of such a photon initiated cascade in the Earth's magnetic 
field by applying the spectra in the relativistic quantum domain 
for the $e^\pm$ pair production by photons and for the
emission of synchrotron photons by secondary pairs given by Baring~(1988). 
The spectra of $e^\pm$ pairs and synchroton photons computed using these formulae
start to differ significantly in their high energy end in respect to the spectra 
without quantum corrections (e.g. Erber  1966) if the parameters $\chi_\gamma$ and 
$\chi_{e^\pm}$ are much higher than one.
In order to show the quantitative differences between the spectra
with and without quantum corrections, Fig.~\ref{fig0} presents the probabilities of 
production of $e^\pm$ pairs by photons and the probabilities of emission of 
synchrotron photons by electrons (positrons) for selected values of  
$\chi_\gamma$ and $\chi_{e^\pm}$. These probabilities have been calculated from
\begin{eqnarray}
P = \lambda_{e,\gamma}\int_0^{\varepsilon}R_{e,\gamma}d\epsilon, 
\label{eq1}
\end{eqnarray}
\noindent
where $P$ is the random number. $\varepsilon$ is the ratio of a simulated energy
of the electron (positron) normalized to the energy of the parent photon
for the process of photon conversion into an $e^\pm$ pair, or 
the ratio of simulated energy
of synchrotron photon normalized to the energy of the parent electron (positron)
for the process of synchrotron emission of photon by an electron (positron).
$R_\gamma$ and $R_e$ are the spectra for production of $e^\pm$ pairs by the photon
or for production of synchrotron photons by the electron (positron) (see Baring~1988). 
The mean free paths, $\lambda_{e,\gamma}$, for these processes are obtained from
\begin{eqnarray}
\lambda^{-1}_{e,\gamma} = \int_0^{1}R_{e,\gamma}d\epsilon.
\label{eq2}
\end{eqnarray}
\noindent
From Fig.~\ref{fig0}, it is evident that if the values of $\chi >> 1$, 
the energies of produced secondary particles are significantly higher when the 
quantum corrections are included. 
Therefore, the more precise formulae given by Baring should be used if the
parameters $\chi_\gamma$  and $\chi_{e^\pm}$ are of the order of a few.  
This is the case of photons with the parameters of some highest energy AGASA 
events ($> 10^{20}$ eV). With these exact formulae the cascade initiated by photons 
becomes more penetrating.  

\begin{figure} 
  \vspace{5.cm} 
 \includegraphics{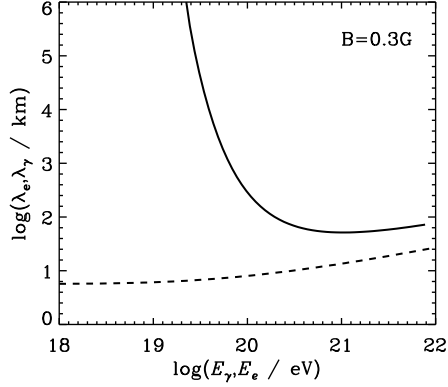}
\caption[]{The dependence of the mean free paths for photon conversion with 
energy $E_\gamma$ into an $e^\pm$ pair (full curve) and for production of the  
synchrotron photon by an electron (positron) with energy $E_{e^\pm}$
in the perpendicular magnetic field with the strength of 0.3 G
on energy of the primary particle.}      
\label{fig1}  
\end{figure}

To find out which specific 
photons can initiate cascades in the Earth's magnetic field,   
we have computed the mean free path for a conversion of photons
with energies $E_\gamma$ into $e^\pm$ pairs in a perpendicular magnetic field
and for the production of synchrotron photons by
electrons (positrons) with energies $E_e$ in this same magnetic field.
In Fig.~\ref{fig1} we show the results as a function of photon and electron 
energies in a perpendicular magnetic field with the strength of 0.3 G 
(typical of the Earth's surface on the equator).
The mean free paths for other magnetic fields can be obtained 
from Fig.~\ref{fig1} by simple linear scaling of the present computations 
(shift to the left and down for a stronger magnetic field). 
From Fig.~\ref{fig1} it becomes clear that photons with the energies
above $\sim 3\times 10^{19}$ eV may convert into $e^\pm$ pairs
(the mean free path comparable to the Earth's radius). The shortest 
mean free paths ($\sim  45$ km) have photons with the energies of a few 
$10^{20}$ eV. The mean free path for the production of synchrotron photon 
by an electron (positron) is equal to $\sim 6$ km for 
$E_{e^\pm} < 10^{19}$ eV. Our computations of the mean free paths for these
two processes are consistent with the results obtained by Kasahara~(1997).

To analyse the cascade initiated by a specific photon with the parameters of the 
AGASA event we apply the Monte Carlo method. We start to follow the propagation of 
a photon from the distance of 5 Earth radii. The distance, $x_\gamma$, at which the 
photon converts into an $e^\pm$ pair is obtained from: 
\begin{eqnarray}
P_2 = exp (-\int_0^{x_\gamma}\lambda_{\gamma}^{-1}dx)
\label{eq3}
\end{eqnarray}
\noindent
where $P_2$ is the random number.  Note that $\lambda_{\gamma}$ 
depends on the propagation distance $x$.
The energies of secondary electrons (positrons) are obtained from Eq.~\ref{eq1}.
The distance, $x_{e^\pm}$, of the emission of a synchroton photon by an electron 
(positron) is obtained from Eq.~\ref{eq3} in which $\lambda_{\gamma}$ is
replaced by the mean free path for emission of synchrotron photon $\lambda_{e}$
(see Eq.~\ref{eq2}). The energy of synchrotron photon produced by secondary electron
(positron) is obtained from Eq. \ref{eq1}. 

Our simulations show that in the case of cascades initiated 
by photons with the energies of $\sim 10^{20}$ eV in the 
magnetic field with a typical value of 0.3 G, up to 4 pairs of secondary $e^\pm$ pairs
can be produced. These 8 leptons produce on their path through the magnetosphere up 
to a few hundred synchrotron photons which arrive in the Earth's atmosphere with 
the energies above 10 GeV. Since leptons produced in the Earth's magnetosphere seperate 
from each other (different charge) and their number is
relatively low, the collective effects on emission of synchrotron radiation
such as backreaction (Nelson \& Wasserman 1991, Aloisio \& Blasi 2002a)  
or coherent emission (Aloisio \& Blasi 2002b) are neglected. 

We conclude that the photons with the parameters of the highest energy events 
observed in cosmic rays can convert into $e^\pm$ pairs in the Earth's 
magnetic field. However, due to the dipole structure of the Earth's magnetic
field, the probability of conversion should strongly depend on the 
location of the cosmic ray observatory and on the arrival direction of the primary
photon.

\section{Photons arriving to the location of the AGASA array.}

\begin{figure} 
  \vspace{5.cm} 
 \includegraphics{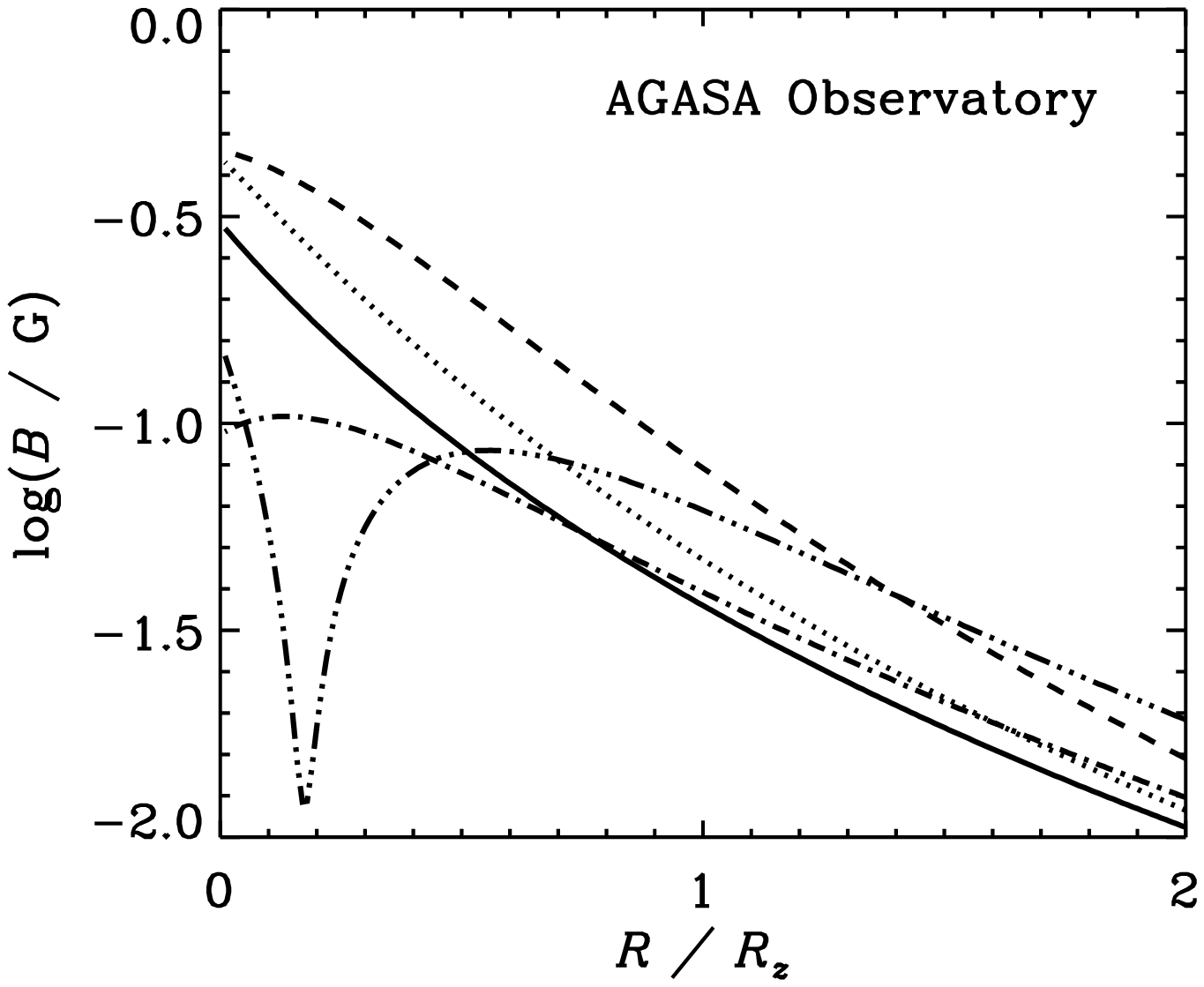}
 \includegraphics{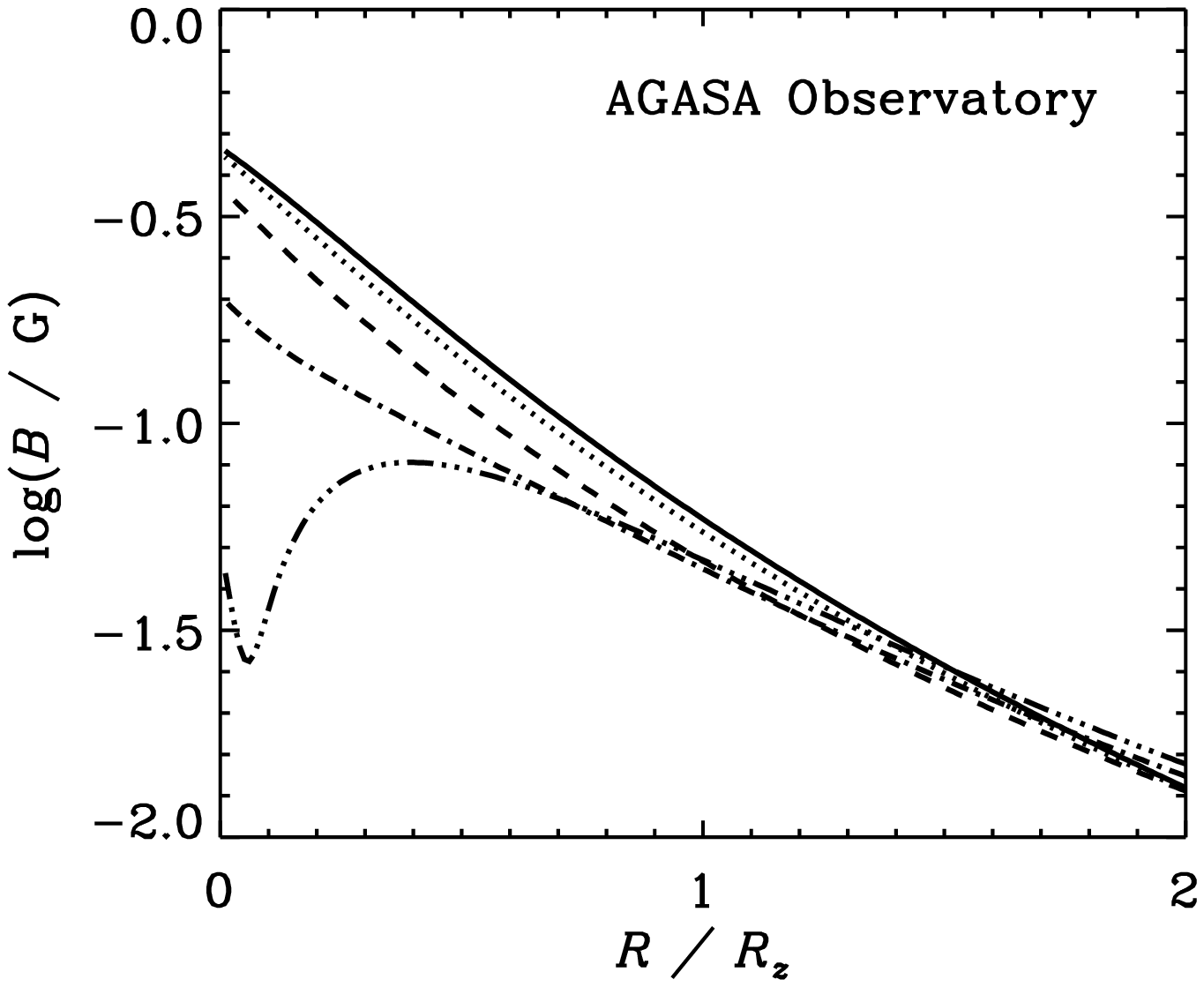}
\caption[]{
The magnetic field profiles along the direction of EHE photons motion  
at the location of the AGASA Observatory (the distance measured in units 
of the radius of the Earth $R_z$). The left figure shows the magnetic 
field profiles as a function of the zenith angle: $Z = 0^{\rm o}$ 
(full); $A = 0^{\rm o}$, and
$Z = 30^{\rm o}$ (dotted), and $Z = 60^{\rm o}$ (dashed); and $A =
180^{\rm o}$, and $Z = 30^{\rm o}$ (dot-dashed), and $Z = 60^{\rm o}$
(dot-dot-dot-dashed), and the right figure for $Z = 45^{\rm o}$ as a function
of azimuth angle;  $A = 0^{\rm o}$ (full), $45^{\rm o}$ (dotted), $90^{\rm o}$
(dashed),  $135^{\rm o}$ (dot-dashed), and $180^{\rm o}$ (dot-dot-dot-dashed).}
      \label{fig2}  
\end{figure}

\begin{figure*}
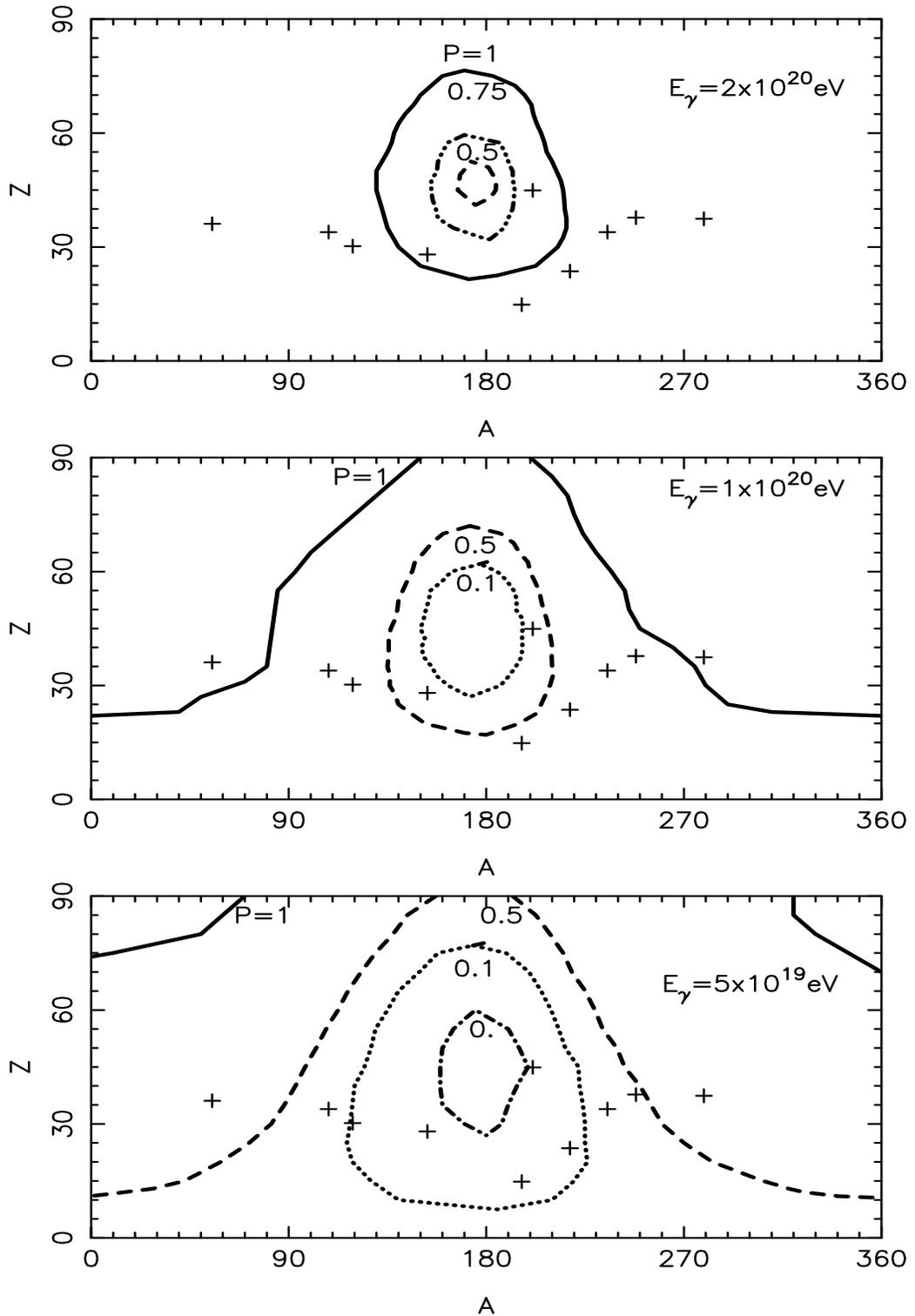
 
  \vspace{22.cm} 
\includegraphics{agasaprawd2e20.eps}     
\includegraphics{agasaprawd1e20.eps}   
\includegraphics{agasaprawd5e19.eps}     
\caption[]{The probability of conversion of
photons with energies $2\times 10^{20}$ eV, $10^{20}$ eV, and $5\times 
10^{19}$ eV, arriving in the AGASA array from different 
directions defined by the zenith
$Z$ and azimuth $A$ angles. The probability, $P$, is less than one inside the
region marked by the full curve, $P < 0.75$ (dot-dot-dot-dashed), $P < 0.5$
(dashed), $P < 0.1$ (dotted), and $P = 0.$ (dot-dashed). The crosses
mark the arrival directions of the AGASA events with energies 
$> 10^{20}$ eV.}
\label{fig3}  
\end{figure*}

In order to have an idea which photons with the highest
energy AGASA events can interact with the Earth's magnetosphere, we compute the 
profiles for the perpendicular components of the magnetic field for 
different photon arrival directions in respect to the location of the AGASA 
observatory ($35^o47'$N, $138^o30'$E), using the World Magnetic 
Model\footnote{available  through: www.ngdc.noaa.gov}. 
The results of the calculations for different 
zenith $Z$ and azimuth $A$ angles (measured clockwise from the North) 
are shown in Figs.~\ref{fig2}. It  is evident that photons arriving to the
AGASA array from the southern directions (the azimuths $90^{\rm o} < A < 
270^{\rm o}$) meet  the lowest perpendicular component of the magnetic 
field along their way of propagation.
For these azimuth angles, and at some range of the zenith angles, the perpendicular
component of the magnetic field can reach very small values since in these
directions of the sky the magnetic field is almost tangent to the path of the photon 
(see characteristic cusps in Figs.~\ref{fig2}). The magnetic field profiles
determine the probability of conversion of a primary photon with
specific energy into an $e^\pm$ pair. In Figs.~\ref{fig3} we show the
probabilities for interaction of photons with
different energies: $E_\gamma =2\times 10^{20}$ eV, $10^{20}$ eV, and $5\times 10^{19}$ 
eV, coming to the AGASA array from the sky. These probabilities
strongly depend on the arrival directions and photon energies.
Photons with the energies $> 10^{20}$ eV, arriving from the northern sky for
zenith angles greater than $\sim 20^{\rm o}$, should convert into $e^\pm$ pairs with
the probability equal to one. However, for the southern sky, there is a region 
centered on $Z\sim 45^{\rm o}$ and
$A\sim 175^{\rm o}$ (slightly shifted from the South to the East due to the 
south-east Asian magnetic anomaly), where the probability of interaction is 
significantly lower than one even for the highest energy AGASA events. The
probability of photon interaction drops very fast
with photon energy. For example, photons with the energies of $5\times 10^{19}$ eV 
already have a chance
to arrive in the Earth's atmosphere without interaction even from the 
northern sky. For such photons the region of low interaction 
probability expends largely containing in its center the region with
the probability of interaction equal to zero.

\subsection{Cascading effects of photons with the parameters of the 
AGASA events.} 

\begin{figure} 
  \vspace{5.cm} 
 \includegraphics{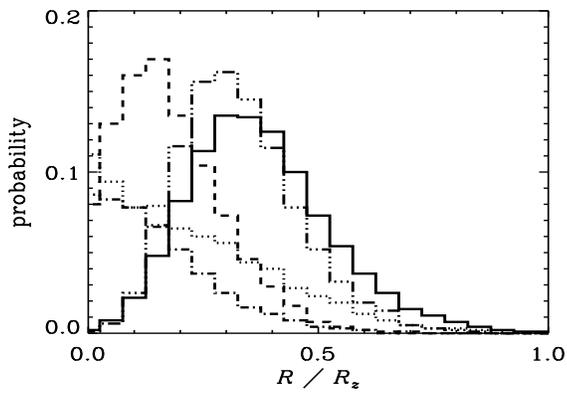}
\caption[]{The probability $(\Delta {\rm N}/\Delta {\rm log (R/R_Z))}$ of the 
first interaction of photons with parameters of some AGASA events 
($2.13\times 10^{20}$ eV - full histogram, $1.5\times 10^{20}$ eV -
dotted, $1.2\times 10^{20}$ eV - dot-dashed, $1.04\times 10^{20}$ eV -
dot-dot-dot-dashed, and $1.01\times 10^{20}$ eV - dashed) as a function 
of distance from the surface of the Earth, measured in units of the Earth's
radius $R_Z$. The histograms are averaged over $10^4$  simulated primary
photons.}   
\label{fig4}  
\end{figure}

Let us assume that the highest energy AGASA events are caused by the EHE photons.
For photons with the energies of some of these events the parameter $\chi_\gamma$
can be as high as a few. Therefore these photons should  
pre-cascade in the Earth's magnetosphere before developing the cascade in the 
Earth's atmosphere. Published
parameters of the AGASA events with the energies above $5\times 10^{19}$
eV (Hayashida et al. 2000) allows us to re-calculate their arrival directions, i.e.
the zenith and azimuth angles (see  Table~\ref{table1}). Using them, we simulate 
cascades initiated by photons in the 
Earth's magnetosphere. The probabilities that photons will cascade with
the parameters of these events (averaged over $10^4$  simulations) and the numbers 
of secondary photons with energies $> 10^{17}$ eV (averaged over 200 
simulations) arriving at the  distance of 20 km from the Earth's  
surface are reported in the fifth and the sixth column of Table~\ref{table1}.
First photon interaction usually occurs closer to the detector than the one 
Earth's radius. The most probable distance of the first interaction is usually closer 
than $\sim 2500$ km from  the AGASA array (the results for some specific events 
are shown in Fig.~\ref{fig4}).

It is interesting to note that most of the AGASA events with the energies $> 10^{20}$
eV (eight out of ten) tend to arrive from the southern part of the sky (their 
directions 
are marked by crosses in Figs.~\ref{fig3}), i.e from the part of the sky where 
the probabilities of a photon conversion into an $e^\pm$ pair are the lowest.
However, our simulations show that the photons
with the parameters of these events cascade efficiently in the Earth's
magnetic field. The probabilities of photon conversion into an $e^\pm$ pair 
are very high for most events with energies $> 10^{20}$ eV (see Table~\ref{table1}).
The photons with the parameters of three AGASA events, which have energies in the range
$(9 - 10)\times 10^{19}$ eV, seem to behave similarly to the photons with energies
$> 10^{20}$ eV. These two lower energy events, which arrive from the southern 
directions, also have a high probability of cascading. 

The situation with lower energy events, i.e. the events with the energies of 
$(5 - 8)\times 10^{19}$ eV, is different. Their arrival directions
are quite isotropic in the azimuth since nine out of twenty events arrive from 
the southern sky and the rest from the northern sky. Only the events 
arriving from the northern sky cascade with a high probability
(see Table~\ref{table1}).  For example, two AGASA events with the same energies
$5.53\times 10^{19}$ eV (91/05/31 and 92/02/01), arriving from the
northern and southern directions cascade 
with a completely different probability (equal to  0.935 and 0.005,
respectively). Therefore, basing on the features of the cascades initiated by the 
photons with parameters 
of the AGASA events we distinguish two groups of particles. The first one contains 
events with the energies above $9\times 10^{19}$ eV (13 events) which tend to arrive 
from the southern directions. The second one contains 
events with the energies below $8\times 10^{19}$ eV (20 events). Their distribution of
arrival directions in the azimuth is more uniform. These two groups are separated
by the lack of events with the energies in the range $(8 - 9)\times 10^{19}$ eV.

In Table~\ref{table1} we also report the number of secondary photons with 
the energies $> 10^{17}$ eV produced
in cascades initiated by these events. These secondary photons carring most of
the energy of a primary photon, fall into the Earth's atmosphere within
a very small cone with the radius of a few tens of centimeters. They develop
electromagnetic showers which in fact cannot be distinguished by cosmic ray detectors
from a single shower. The secondary electrons and positrons
produced in the cascade (usually not more than eight particles)
arrive in the Earth's atmosphere with typical energies above
$\sim 10^{18}$ eV. They carry only a few percent
of the primary photon energy so their contribution to the cascade in the
atmosphere is negligible.  

\begin{figure}[t] 
  \vspace{21.cm} 
 \includegraphics{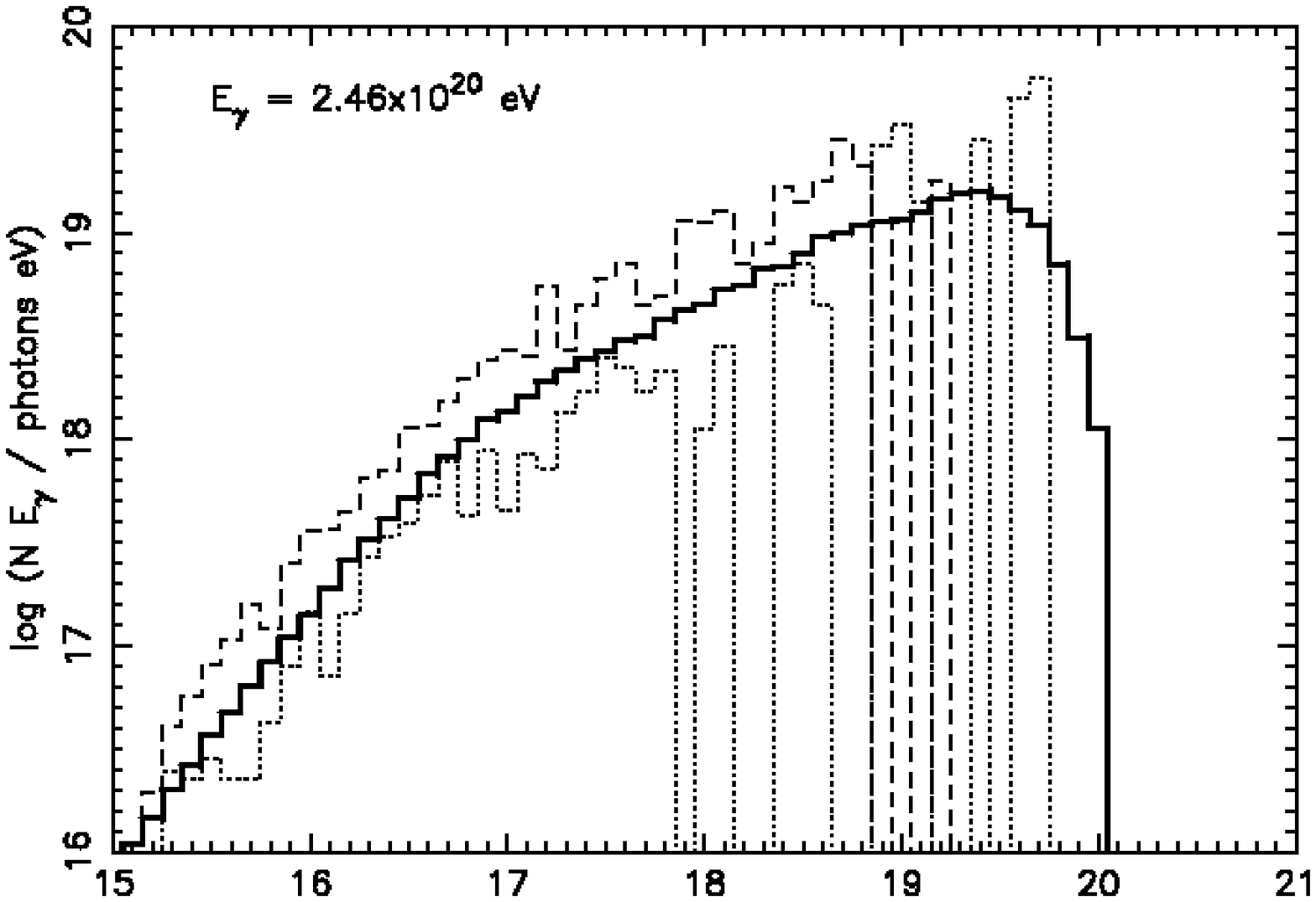}
 \includegraphics{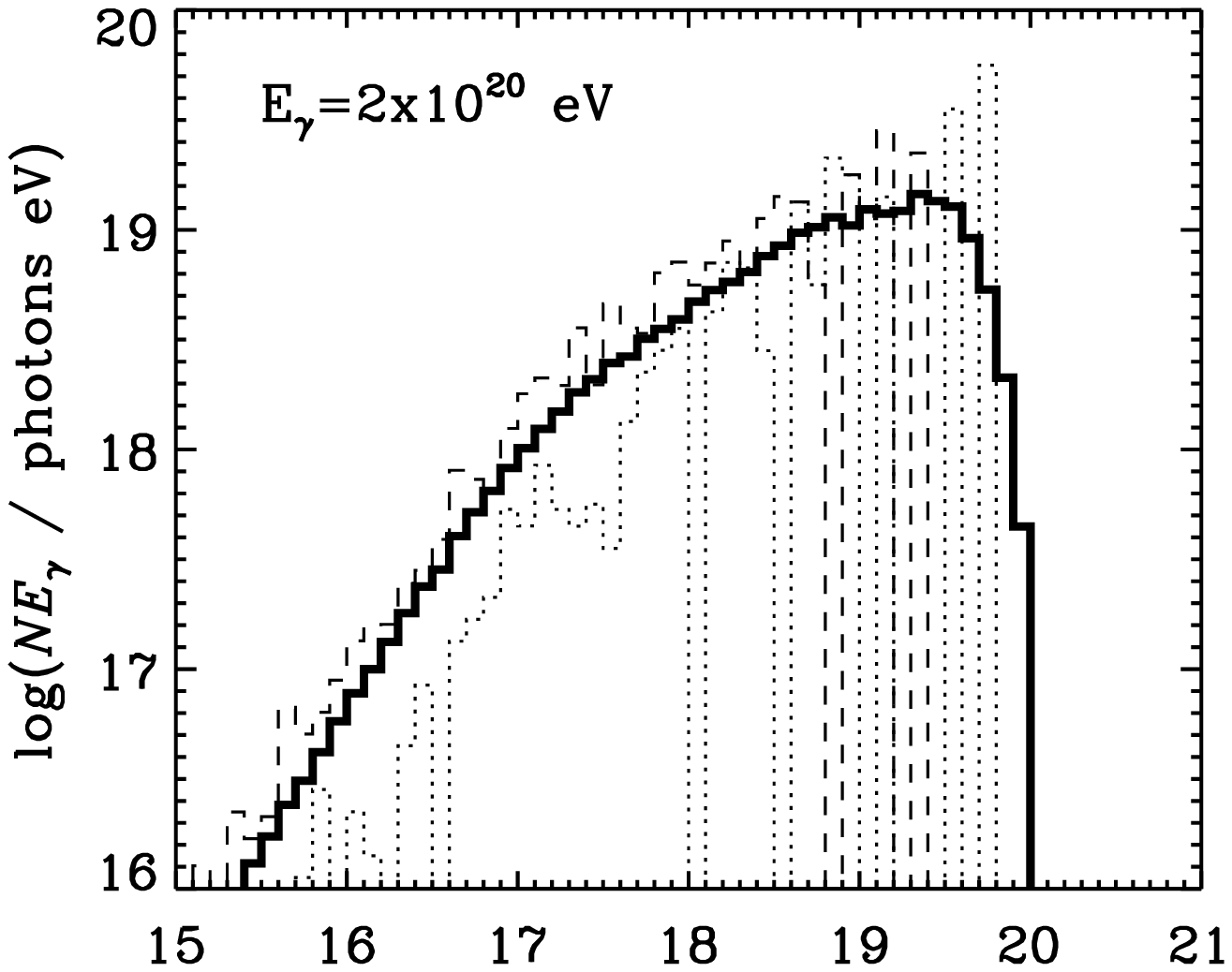}
 \includegraphics{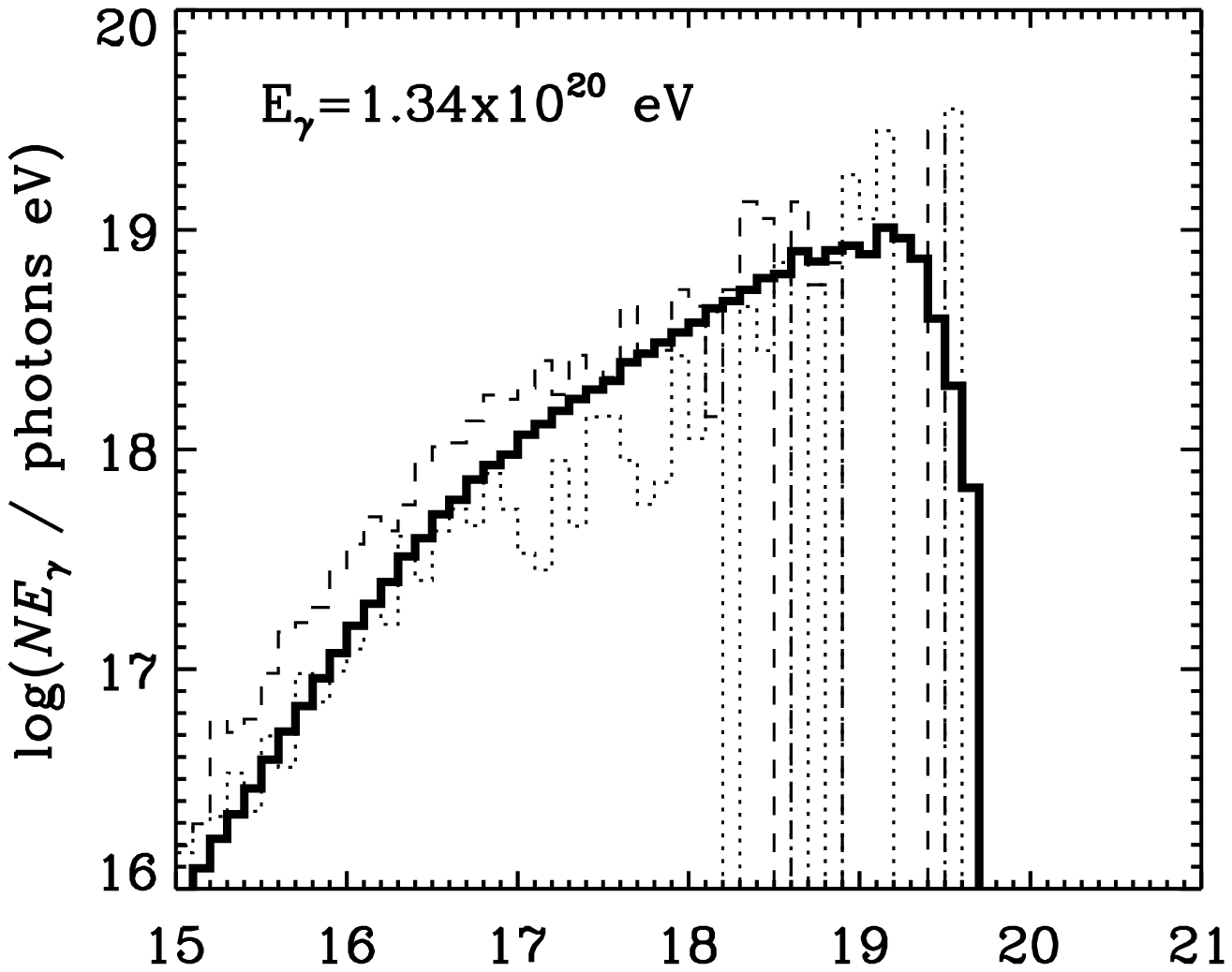}
\includegraphics{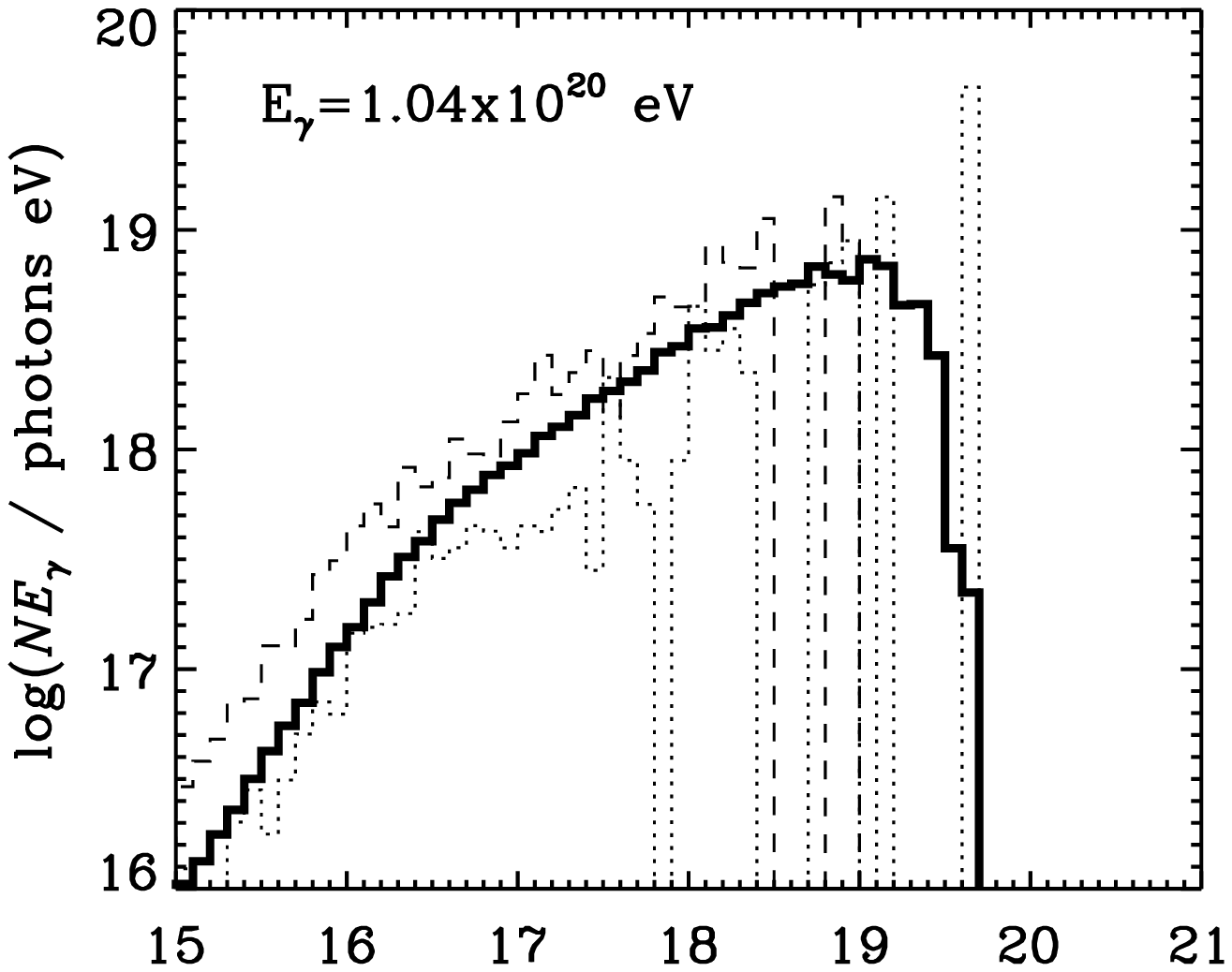}
 \includegraphics{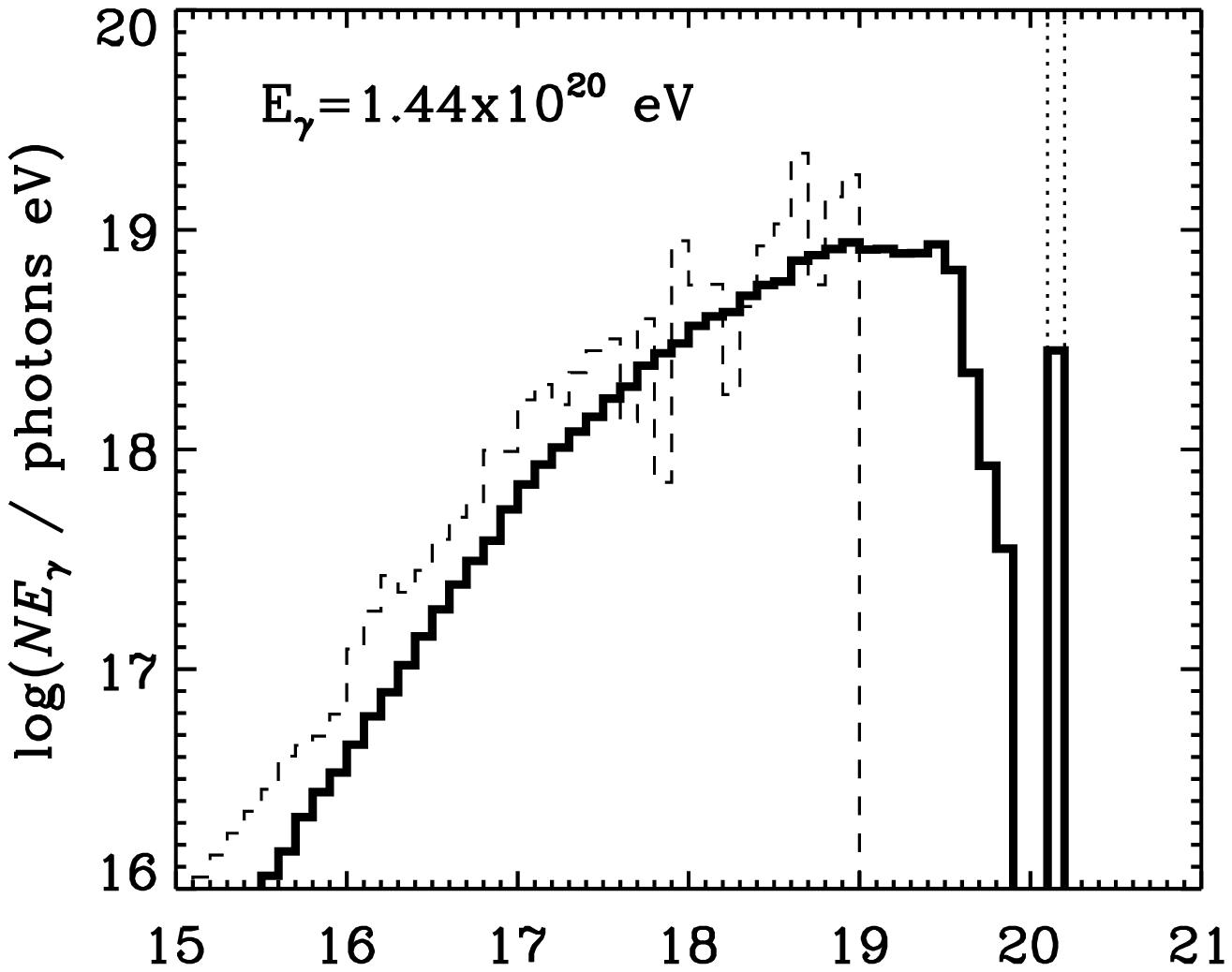}
\includegraphics{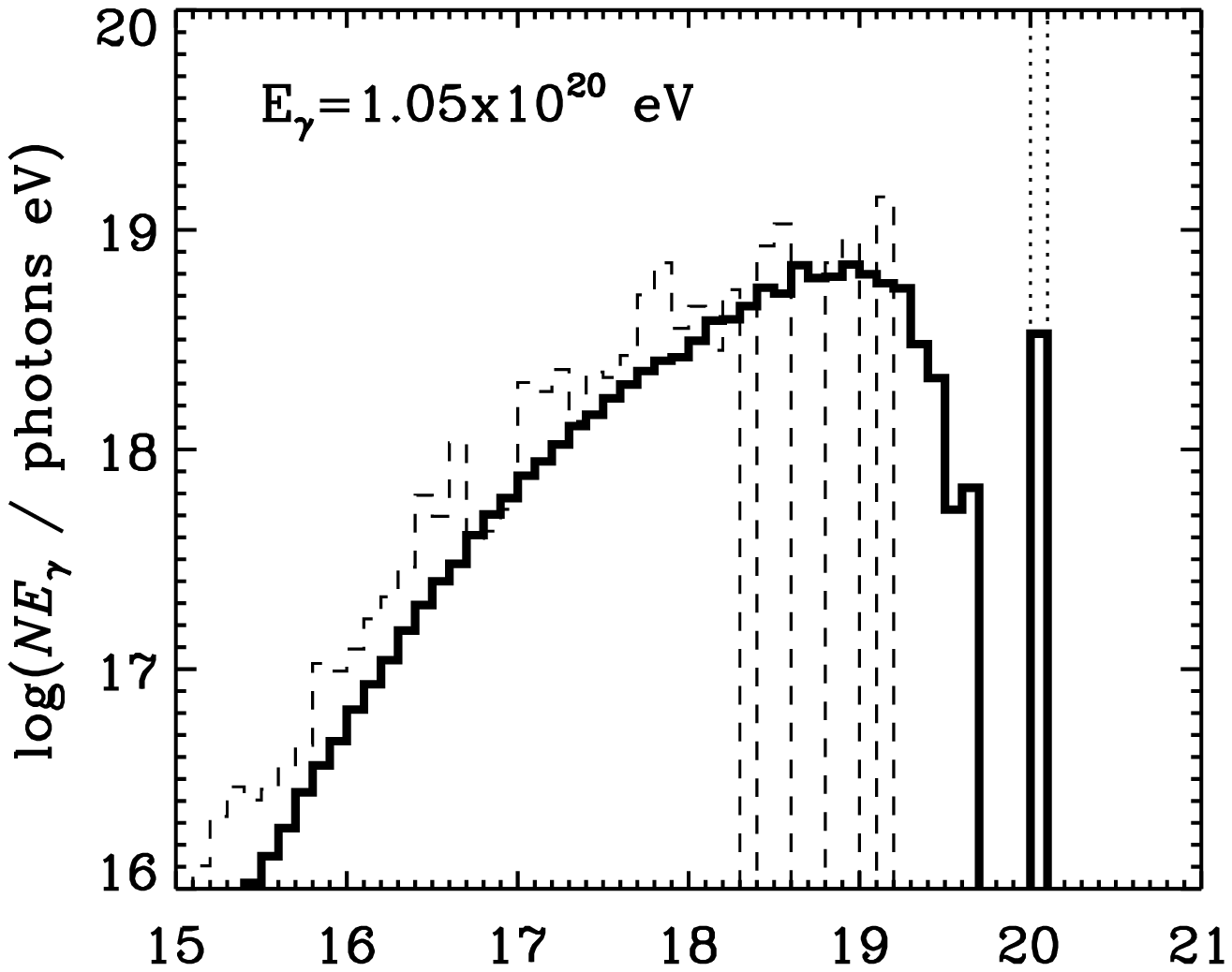}
 \includegraphics{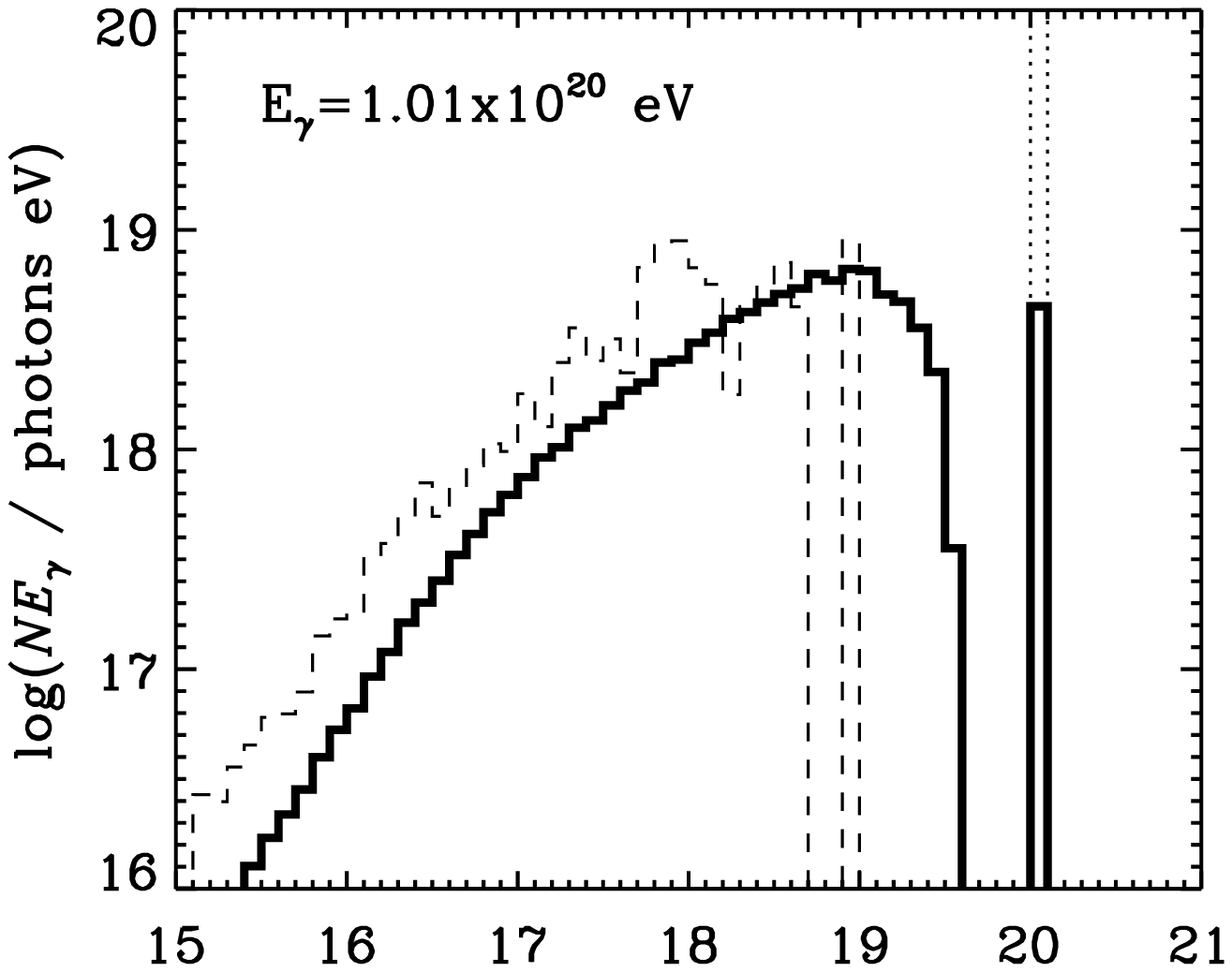}
 \includegraphics{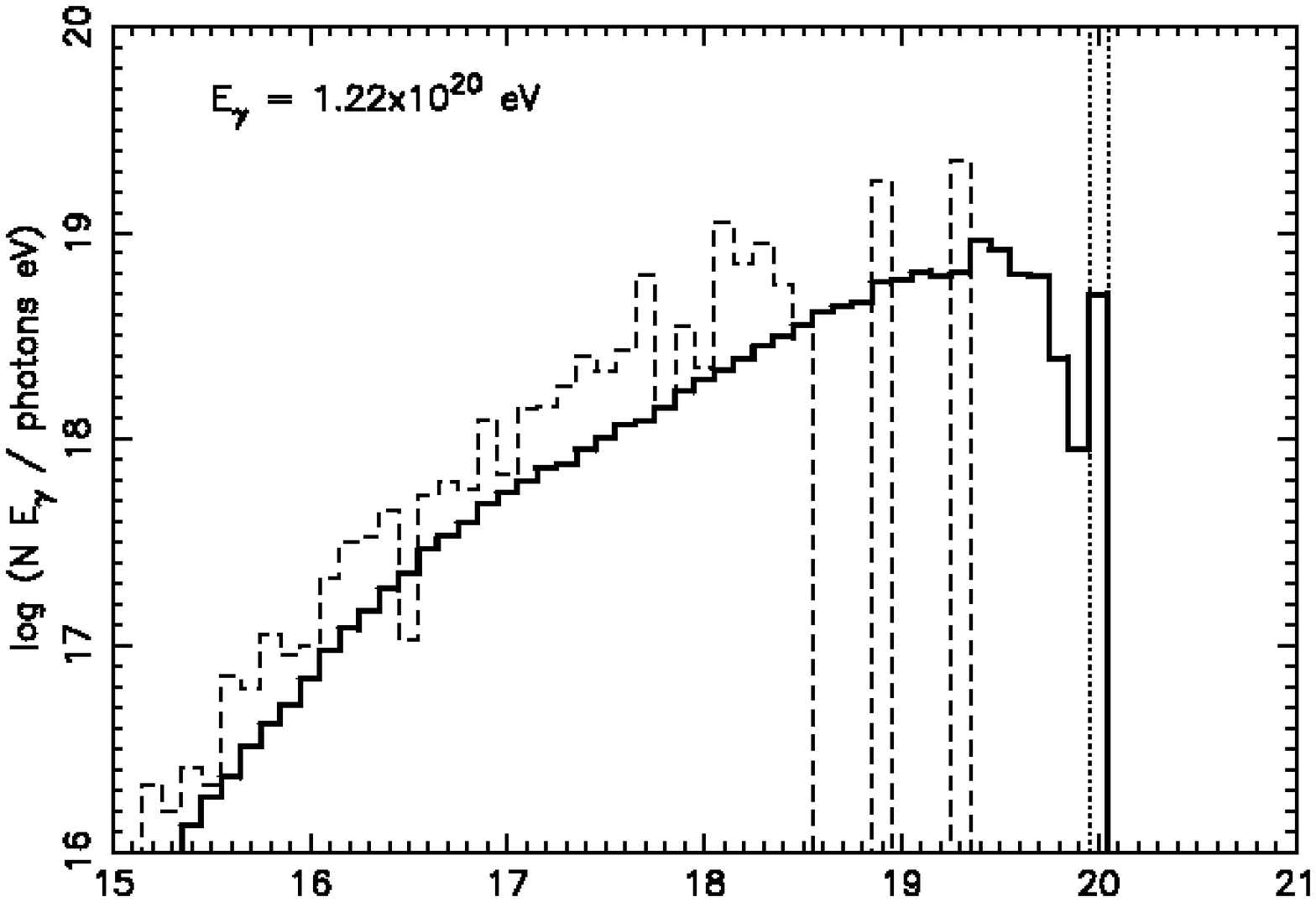}
\includegraphics{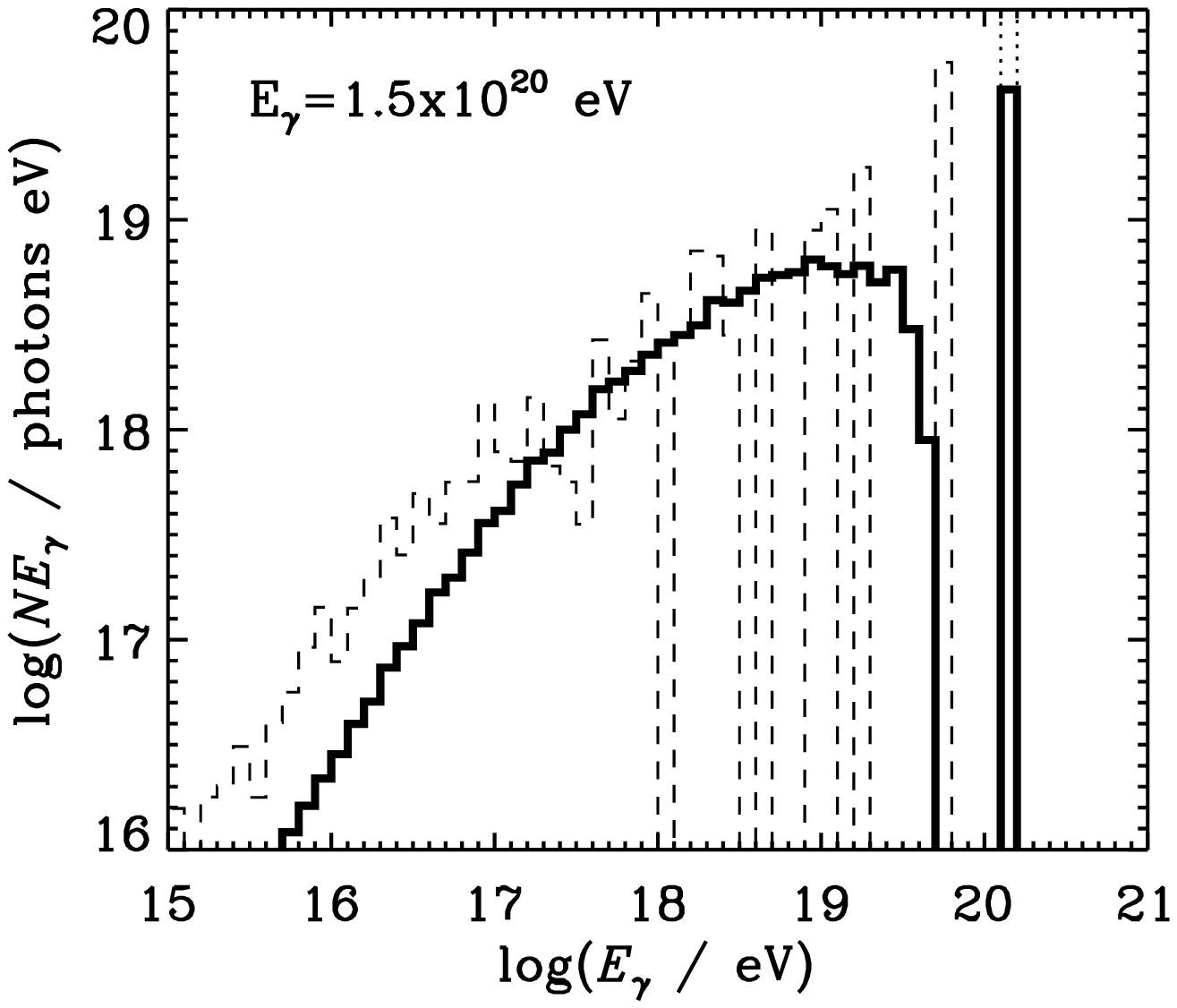}
 \includegraphics{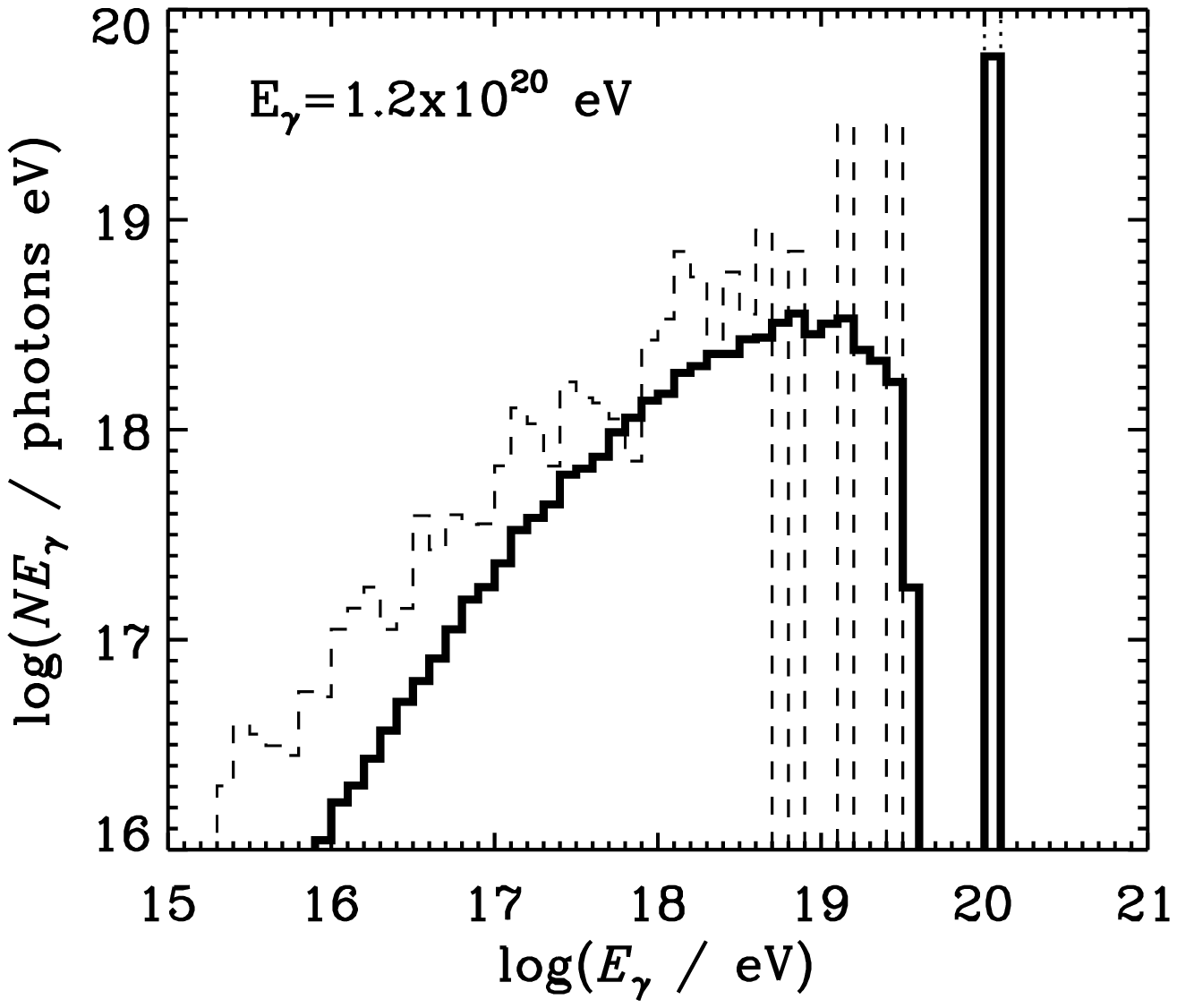}

\caption[]{Average energy distributions of secondary photons (full histograms)
coming in the Earth's atmosphere. These secondary photons are produced in cascades
initiated by the primary photons with the parameters of
the AGASA events with the energies $> 10^{20}$ eV (for other parameters see 
Table~1). The dotted and dashed histograms show the simulations with the smallest
and the largest numbers of secondary photons selected from 200 simulated primary
photons.}   
\label{fig5}  
\end{figure}

We also investigate the spectra of secondary cascade photons, produced in 
the magnetic field by primary photons with the parameters of the AGASA events with
the energies $> 10^{20}$ eV, and fluctuations of these spectra in respect to
the spectrum averaged over 200 simulations.  In Fig.~\ref{fig5} 
we show the histograms with the numbers of  secondary photons within the 
specific energy range ($N = \Delta N_\gamma / \Delta (log E_\gamma)$, where
$\Delta (log E_\gamma) = 0.1$),
which arrive in the atmosphere (i.e. at the distance of 20 km from the
surface). The results of the simulations with the lowest and the largest numbers of 
secondary  photons (selected from these 200 simulations) are marked by the 
dotted and dashed
histograms. In the case of four AGASA events (with energies: $2.46\times
10^{20}$ eV, $2.13\times
10^{20}$ eV, $1.34\times 10^{20}$ eV, and $1.04\times 10^{20}$ eV) all 200
simulated photons converted into $e^\pm$ pairs. The fluctuations 
between different simulations in the 
number of secondary photons with energies $> 10^{15}$ eV may be as high as 
a factor of 2-3 around the average values. However, these 
fluctuations mainly concern the
lower energy secondary photons and therefore, do not affect 
the development of cascades in the atmosphere seriously. The fluctuations in the
location of the maximum in the spectrum are more important since 
this maximum is usually located close to or just above $\sim 10^{19}$ eV. 
Above this energy
the Landau-Pomeranchuk-Migdal (LPM) (Landau \& Pomeranchuk~1935, Migdal 1956)
effect can have a significant influence on their further cascading in the 
atmosphere. The fluctuations of the location of the maximum in the spectrum of 
secondary photons strongly depend on the considered AGASA
event.  They are only of the order of two in the case of events with the energies
of $2.13\times 10^{20}$ eV and $1.34\times 10^{20}$ eV. However, they can
be as high as a factor of four for the event with the energy of $1.04\times 10^{20}$ eV.   
In fact, the primary photons with the parameters of some events with the energies 
$> 10^{20}$ eV may not cascade at all in the magnetosphere. 
This possibility takes place 
rather rare, since all these events, except the event with the energy of $1.2\times
10^{20}$ eV, cascade with a probability greater than a half. 

A closer inspection of Fig~\ref{fig5} shows that most
primary photons with parameters of the AGASA events with the energies $> 10^{20}$
eV arrive in the atmosphere as a bunch of lower energy secondary photons with
the mean energy just above $10^{19}$ eV. Therefore, these primary photons should 
initiate cascades in the Earth's atmosphere which are not influenced by
the LPM effect. The LPM effect is
responsible for large fluctuations of showers which are
induced by photons with the energies above a few $10^{19}$ eV, as recently
verified in the Monte Carlo simulations based on the Cosmos code
(Kasahara~1997)
and AIRES program (Bertou, Billoir \& Dagoret-Campagne 2000). Also
the depth of the  maximum of showers simulated with the LPM  effect is much
deeper in the atmosphere than in the case of showers without the LPM effect.
That is why it is generally expected that showers in the atmosphere
initiated by the EHE photons with the energies above a few $10^{19}$ eV should be
easily distinguished from the ones induced by hadrons. However, as we noted
above, primary photons with parameters of the AGASA events with energies 
above $>10^{20}$ eV usually pre-cascade in the magnetosphere with a high
probability. Their products arrive in the atmosphere in the form of bunches of 
secondary photons initiating cascades in the atmosphere which are not significantly
influenced by the LPM effect.
The depth of the shower maximum which is induced by photons with the energies  of 
$\sim 10^{19}$ eV (equal to $\sim 900$ g cm$^2$) does not differ significantly from
the depth of shower maximum induced by the protons with energies of
AGASA  events (Bertou, Billoir \& Dagoret-Campagne 2000). 

The photons with parameters of the lower energy AGASA
events $< 8\times 10^{19}$ eV, which arrive from the southern directions, 
do not
cascade in the  Earth's magnetosphere or cascade with a low probability
(Table~\ref{table1}).  The cascades initiated in the atmosphere by these
primary photons, which do not suffer pre-cascading in the magnetosphere, are
susceptible to large fluctuations due to the LPM effect. Hence, these
showers can have the shower maximum even deeper in the atmosphere than showers
initiated by primary photons with significantly higher energies which
pre-cascaded in the magnetosphere.

\subsection{Cascading effects of photons with reduced energies of 
the AGASA events}

The conclusion mensioned in the previous section 
suggests that also the photons which have the arrival 
directions of AGASA events with energies $> 10^{20}$ eV, but  
energies reduced by a factor of two or three in respect to the values estimated 
by AGASA group, should arrive in the atmosphere without pre-cascading 
in the magnetosphere. To check this we simulate the cascades initiated by 
primary photons with the parameters of the highest energy AGASA events but with 
energies reduced by factors mentioned above. It is found that in most cases the
probability of conversion of such photons becomes significantly lower (only 
two events, 01/05/10 and  94/07/06, have the probability of interaction still greater 
than a half, see columns seven and eight in Table~\ref{table1}). 

\begin{figure}[t] 
  \vspace{5.cm} 
\includegraphics{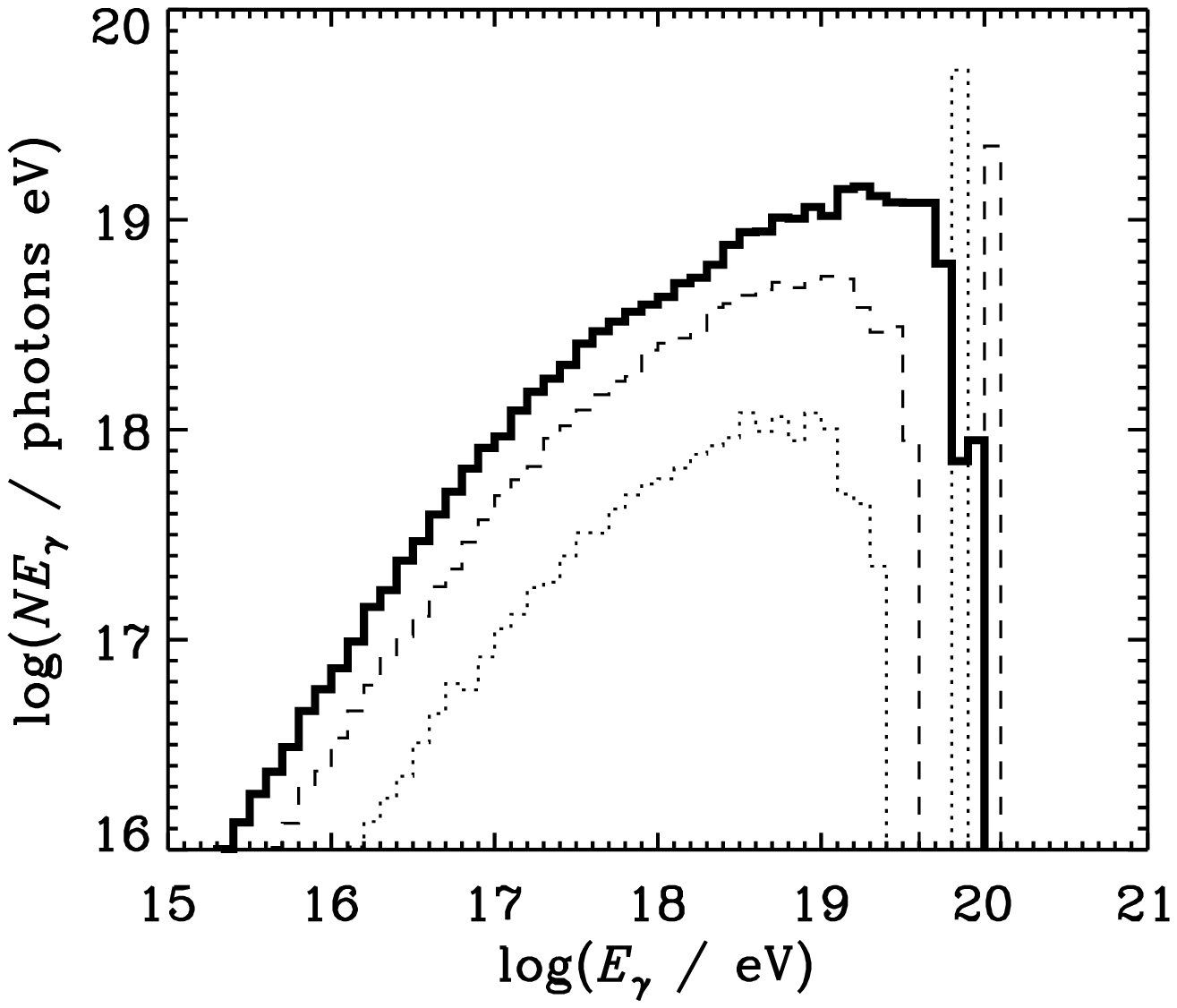}
\includegraphics{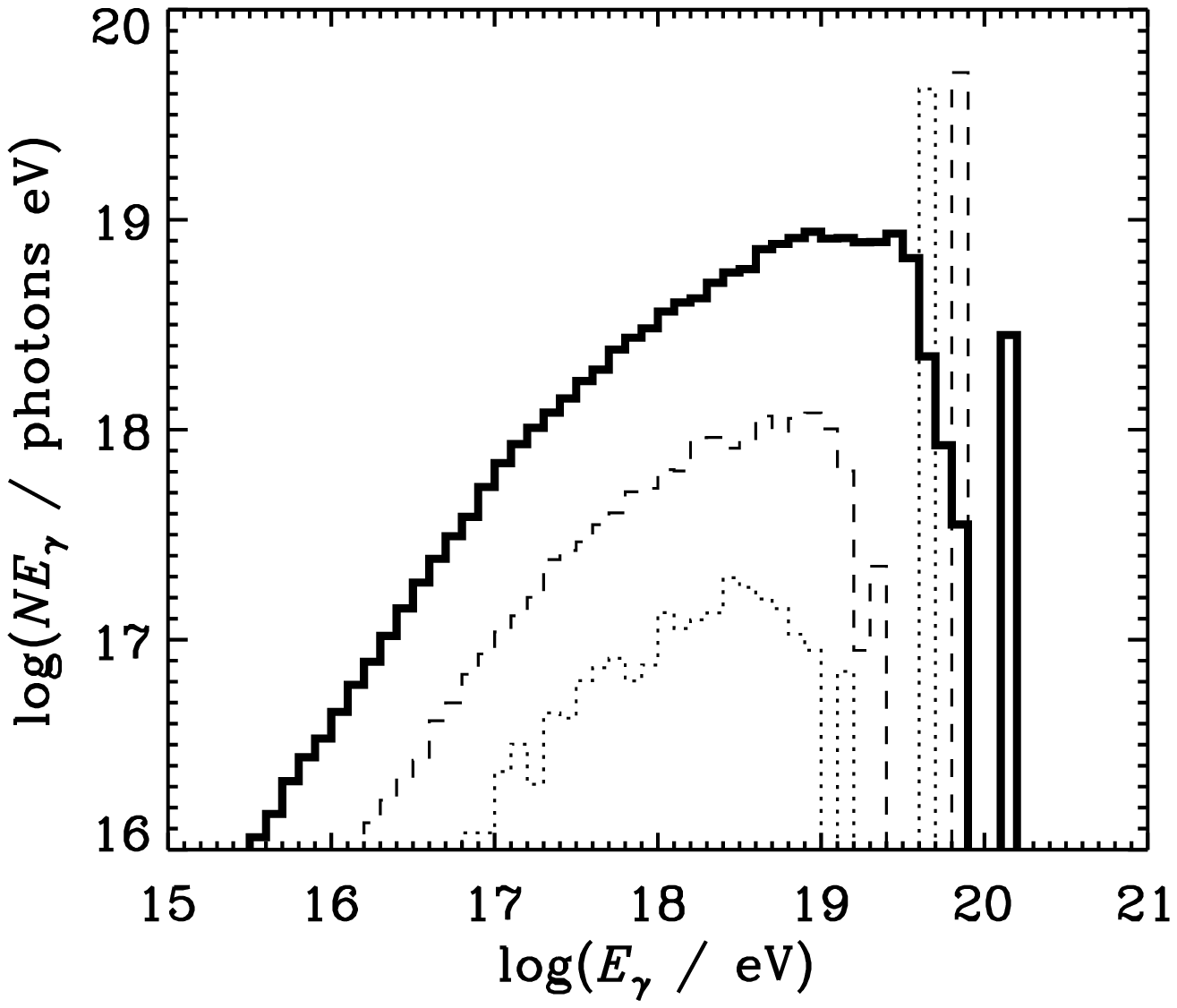}  
\caption[]{Average energy distributions of secondary photons (full histograms),
produced by primary photons with parameters of
the AGASA events 93/12/03, $E_\gamma =2.13\times 10^{20}$ eV (on the left), 
and 96/01/11, $E_\gamma =1.44\times 10^{20}$ eV (on the right), 
averaged over 200 simulated primary photons. The dashed 
and dotted histograms show the cases with the arrival directions 
of these events but energies reduced by a factor of 2 and 3, respectively.}       
\label{fig6}  
\end{figure}

As an example, in Fig.~\ref{fig6} we compare the average 
spectra of secondary photons produced by primary photons with reduced
energies with the ones produced by primary photons with the energies reported by
the AGASA experiment for two events: 93/12/03 and 96/01/11.
In both cases the average energies of secondary photons
are higher in the case of primary photons with reduced energies 
than in the case of the average energies of secondary photons 
produced by primary photons with energies originally estimated by AGASA group 
($2.13\times 10^{20}$ eV  and $1.44\times 10^{20}$ eV). 
Therefore we conclude that the highest energy AGASA events with 
energies reduced by a factor of two or three are more sensitive to 
fluctuations during cascading in the
atmosphere because of a larger influence of the LPM effect on the 
development of showers initiated by the secondary 
photons with higher energies. We suggest that due to these fluctuations,
not taken into account in derivation of the initial energies of
the AGASA events, the energies of AGASA events 
arriving from the southern directions may be incorectly estimated.   

\section{Conclusions}

We analyse the AGASA events with energies above $5\times 10^{19}$ eV for which
the arrival parameters have been published (Hayashida et al.~2000 and AGASA web page). 
It seems that two groups of events can be distinguished. The first group contains 
13 events with energies $>9\times 10^{19}$ eV. 
These events tend to arrive in the AGASA array
mainly from the southern directions (10 out of 13). 
The second group contains 20 events with energies $<8\times 10^{19}$ eV. 
Their arrival distribution in the azimuth angle is uniform (9 from the southern part 
of the sky and 11 from the northern sky).  
These two groups are separated by the lack of events with
energies between $(8 - 9)\times 10^{19}$ eV.

We have shown that primary photons with the parameters of the highest energy AGASA 
events ($> 9\times 10^{19}$ eV) usually efficiently pre-cascade in the Earth's
magnetosphere in spite of the fact that most of them arrive from the 
southern directions where the perpendicular component of the magnetic 
field is the lowest. Photons with parameters of the lower energy AGASA
events, $(5 - 8)\times 10^{19}$ eV, can cascade efficiently when arriving from the 
northern directions. They usually do not cascade (or cascade with a low
probability) when arriving from the southern directions.   

As we have mentioned the primary photons with the energies of $5 - 8\times 10^{19}$ eV
arriving from the southern direction should not cascade 
(or cascade with a low probability). They fall onto the atmosphere with average
energies higher than the average energy of the secondary photons produced in cascades 
initiated in the Earth's magnetosphere by the primary photons with the energies 
$> 9\times
10^{19}$ eV. Therefore, the lower energy primary photons, ($5 - 8)\times 10^{19}$ eV,
should initiate showers in the atmosphere which fluctuate stronger due to the LPM 
effect than the showers initiated by the primary photons with the higher
energies ($> 9\times 10^{19}$ eV). We suggest that these effects may influence 
the estimation of energies of photonic events 
arriving from the southern direction. This suggestion is consistent
with the tendency that the higher energy AGASA events ($>9\times 10^{19}$ eV) 
more likely arrive from the southern directions which is not the case in 
the lower energy events ($< 8\times 10^{19}$ eV). Their arrival 
distribution in the azimuth angle is consistent with the uniformity.

\section*{Acknowledgment.}
\noindent
I would like to thank the anonymous referee for useful comments.
This work is supported by the KBN grant No. 5P03D 025 21.

\newpage

\begin{table}[t]
\begin{center}
\caption{The probabilities of photon interaction.   
\label{table1}}
\begin{tabular}{|l|l|l|l|l|l|l|l|}
\hline 
shower  & $E_\gamma$ & Z & $A$ & P($E_\gamma$) & $N_\gamma(>10^{17}$eV) 
& P($E_\gamma/2$) & P($E_\gamma/3$)\\
\hline      
01/05/10 & 2.46 & $37^{\rm o}$ &  $248^{\rm o}$  &1.00  & ~~~~133.2 &
1.00 & 0.975\\
93/12/03 & 2.13 & $22.9^{\rm o}$ &  $218^{\rm o}$  &1.00  & ~~~~~99.0 &
0.80 & 0.27\\
94/07/06 & 1.34 & $35.4^{\rm o}$ &  $55^{\rm o}$   &1.00  & ~~~~~89.8 &
0.975 & 0.68\\  
99/09/22 & 1.04 & $36.7^{\rm o}$ &  $279^{\rm o}$  &1.00  & ~~~~~76.6 &
0.715 & 0.30\\    
96/01/11 & 1.44 & $14.1^{\rm o}$ &  $196^{\rm o}$  &0.98  & ~~~~~71.2 &
0.29 & 0.06\\ 
96/10/22 & 1.05 & $33.2^{\rm o}$ &  $108^{\rm o}$  &0.97  & ~~~~~69.5 &
0.185 & 0.035\\  
93/01/12 & 1.01 & $33.2^{\rm o}$ &  $235^{\rm o}$  &0.96  & ~~~~~65.7 &
0.165 & 0.02\\     
01/04/30 & 1.22 & $29.5^{\rm o}$ &  $119^{\rm o}$  &0.945  & ~~~~~42.0 &
0.265 & 0.025\\ 
97/03/30 & 1.50 & $44.2^{\rm o}$ &  $201^{\rm o}$  &0.705 & ~~~~~50.1 &
0.02  & 0.0\\
98/06/12 & 1.20 & $27.3^{\rm o}$ &  $153^{\rm o}$  &0.465 & ~~~~~28.9 &
0.01 & 0.0\\  
\hline
84/12/17 & 0.98 & $30.4^{\rm o}$ &  $64^{\rm o}$   &1.00  & ~~~~~73.4 & &\\ 
92/09/13 & 0.93 & $25.6^{\rm o}$ &  $122^{\rm o}$  &0.65  & ~~~~~33.0 & &\\ 
91/11/29 & 0.91 & $45.0^{\rm o}$ &  $178^{\rm o}$  &0.05  & ~~~~~~1.4  & &\\ 
\hline
95/01/26 & 0.776 & $23.7^{\rm o}$ &  $341^{\rm o}$ &1.00  & ~~~~~65.5 & &\\  
99/07/28 & 0.716 & $39.1^{\rm o}$ &  $48^{\rm o}$  &0.995 & ~~~~~61.6 & &\\  
99/01/22 & 0.753 & $44.7^{\rm o}$ &  $125^{\rm o}$ &0.51  & ~~~~~25.0 & &\\  
96/11/12 & 0.746 & $34.1^{\rm o}$ &  $146^{\rm o}$ &0.065 & ~~~~~~3.8  & &\\  
\hline
98/03/30 & 0.693 & $41.8^{\rm o}$ &  $317^{\rm o}$ &1.00  & ~~~~~62.3 & &\\  
84/12/12 & 0.681 & $31.1^{\rm o}$ &  $74^{\rm o}$  &0.93  & ~~~~~49.5 & &\\  
93/06/12 & 0.649 & $22.7^{\rm o}$ &  $44^{\rm o}$  &0.915 & ~~~~~47.0 & &\\  
98/10/27 & 0.611 & $11.5^{\rm o}$ &  $37^{\rm o}$  &0.795 & ~~~~~40.1 & &\\   
86/10/23 & 0.622 & $30.3^{\rm o}$ &  $231^{\rm o}$ &0.48  & ~~~~~20.8 & &\\  
99/10/20 & 0.619 & $33.5^{\rm o}$ &  $154^{\rm o}$ &0.00  & ~~~~~~1.0  & &\\  
\hline
91/05/31 & 0.553 & $37.5^{\rm o}$ &  $343^{\rm o}$&0.935 & ~~~~~50.2  & &\\  
92/08/01 & 0.550 & $27.6^{\rm o}$ &  $328^{\rm o}$&0.925 & ~~~~~46.1  & &\\  
98/04/04 & 0.535 & $30.2^{\rm o}$ &  $38^{\rm o}$ &0.855 & ~~~~~41.7  & &\\  
96/10/06 & 0.568 & $22.^{\rm o}$ &  $327^{\rm o}$ &0.85  & ~~~~~43.2  & &\\  
91/04/03 & 0.509 & $30.1^{\rm o}$ &  $291^{\rm o}$&0.725 & ~~~~~32.9  & &\\  
86/01/05 & 0.547 & $27.1^{\rm o}$ &  $94^{\rm o}$ &0.375 & ~~~~~14.9  & &\\    
95/10/29 & 0.507 & $28.2^{\rm o}$ &  $243^{\rm o}$ &0.295& ~~~~~12.8  & &\\  
95/04/04 & 0.579 & $11.3^{\rm o}$ &  $115^{\rm o}$ &0.265& ~~~~~~9.7   & &\\  
89/03/14 & 0.527 & $~5.2^{\rm o}$ &  $257^{\rm o}$&0.245 & ~~~~~10.4  & &\\  
92/02/01 & 0.553 & $19.6^{\rm o}$ &$203^{\rm o}$ &0.005  & ~~~~~~2.8   & &\\  
      \hline
\end{tabular}
\end{center}
\end{table}

\newpage
 
%

%

\end{document}